# Learning metal microstructural heterogeneity through spatial mapping of diffraction latent space features

**Mathieu Calvat[1] ✉, Chris Bean[1], Dhruv Anjaria[1], Hyoungryul Park[1], Haoren Wang[2], Kenneth Vecchio[2] & J. C. Stinville[1] ✉**

To leverage advancements in machine learning for metallic materials design and property prediction, it is crucial to develop a data-reduced representation of metal microstructures that surpasses the limitations of current physics-based discrete microstructure descriptors. This need is particularly relevant for metallic materials processed through additive manufacturing, which exhibit complex hierarchical microstructures that cannot be adequately described using the conventional metrics typically applied to wrought materials. Furthermore, capturing the spatial heterogeneity of microstructures at the different scales is necessary within such framework to accurately predict their properties. To address these challenges, we propose the physical spatial mapping of metal diffraction latent space features. This approach integrates (i) point diffraction data encoding via variational autoencoders or contrastive learning and (ii) the physical mapping of the encoded values. Together, these steps offer a method to comprehensively describe metal microstructures. We demonstrate this approach on a wrought and additively manufactured alloy, showing that it effectively encodes microstructural information and enables direct identification of microstructural heterogeneity not directly possible by physics-based models. This data-reduced microstructure representation opens the application of machine learning models in accelerating metallic material design and accurately predicting their properties.

The mechanical properties of metallic materials are fundamentally governed by plasticity and its localization at the microstructural scale. Plastic localization originates from the collective behavior of deformation events that are directly controlled by the structure and heterogeneity of the metal's microstructure. For wrought materials, the influential microstructural features have been established over the last decades and notably include annealing twins, triple junctions, quadrupole points, large grains, twist boundaries and macrozones[1–8]. On the other hand, materials produced by additive manufacturing (AM) are associated with differing microstructural heterogeneities related to the rapid cooling of the liquid melt pool during the AM process. These heterogeneities span multiple scales and include heterogeneous grain shapes[9], a combination of high-angle and low-angle grain boundaries[10], cellular structures[11], and nano-scale precipitates[12]. These heterogeneities significantly influence how plasticity localizes in these materials and consequently their mechanical properties[9,10,12,13].

To accurately predict mechanical properties and accelerate materials design[14–21], it is therefore crucial to capture the full range of heterogeneous microstructural features across multiple scales and understand their collective influence on plasticity and deformation mechanisms[22,23]. Electron Backscatter Diffraction (EBSD), a technique that spatially maps point (local) diffraction data, remains the predominant tool for assessing microstructural heterogeneity across various scales. In conventional EBSD analysis, the diffraction patterns are analyzed based on physical considerations and the Kikuchi band locations are extracted to determine crystal structure and crystallographic orientation[24–26]. Beyond crystallographic orientation, diffraction patterns are also known to contain more information, such as lattice expansion[27], dislocation density[11], unit cell chirality[28], magnetization[29,30] or anti-phase domain[31]. For example, sharpness analysis of Kikuchi patterns (diffraction patterns) provides a qualitative assessment of dislocation density and its distribution[11]. Similarly, cross-correlation techniques applied to Kikuchi patterns allow for the evaluation of lattice expansion and its spatial heterogeneity[27]. Additionally, refined spatial analysis of Kikuchi patterns can inform on geometrically necessary dislocation density[32]. These approaches focus on extracting specific physical information, such as crystallographic orientation, while discarding other potentially valuable data embedded in the diffraction patterns. An example of this conventional approach is shown in Fig. 1A, where crystallographic orientation is determined by analyzing the location of Kikuchi bands within the diffraction pattern[25,26,33]. In contrast, when sharpness analysis is used to assess dislocation density, the focus shifts toward evaluating the diffuseness of the Kikuchi bands rather than their precise location. For each of these analyses, some information from the raw collected data is lost during the process of transitioning from reciprocal space (diffraction space) to physical space, and to present the results in a human-interpretable map. Even when combining multiple physics-based analysis methods, we still fail to fully capture all the relevant information contained in these diffraction patterns, potentially overlooking critical insights into microstructural features. Ultimately, microstructure is described spatially by one or more physical descriptors (stereographic projection of physical directions within lattice, for instance), forming maps that are visually informative by themselves but consist of simplified information. In addition to conventional EBSD analysis, which is based on identifying Kikuchi bands[24–26], or the previously mentioned methods to extract additional information, other methods have been used in the literature to analyze diffraction patterns. These methods include data-driven approaches, such as principal component analysis (PCA)[34], multivariate analysis[35,36], and physics-aware methods, such as virtual aperture[37,38]. These methods produce new representations of microstructures (maps) that may not have a direct physical interpretation as they shift toward direct data analysis.

[1]Materials Science and Engineering Department, The Grainger College of Engineering, University of Illinois Urbana-Champaign, Urbana, IL, USA. [2]Department of NanoEngineering, University of California San Diego, San Diego La Jolla, San Diego, CA, USA. ✉e-mail: mcalvat@illinois.edu; jcstinv@illinois.edu

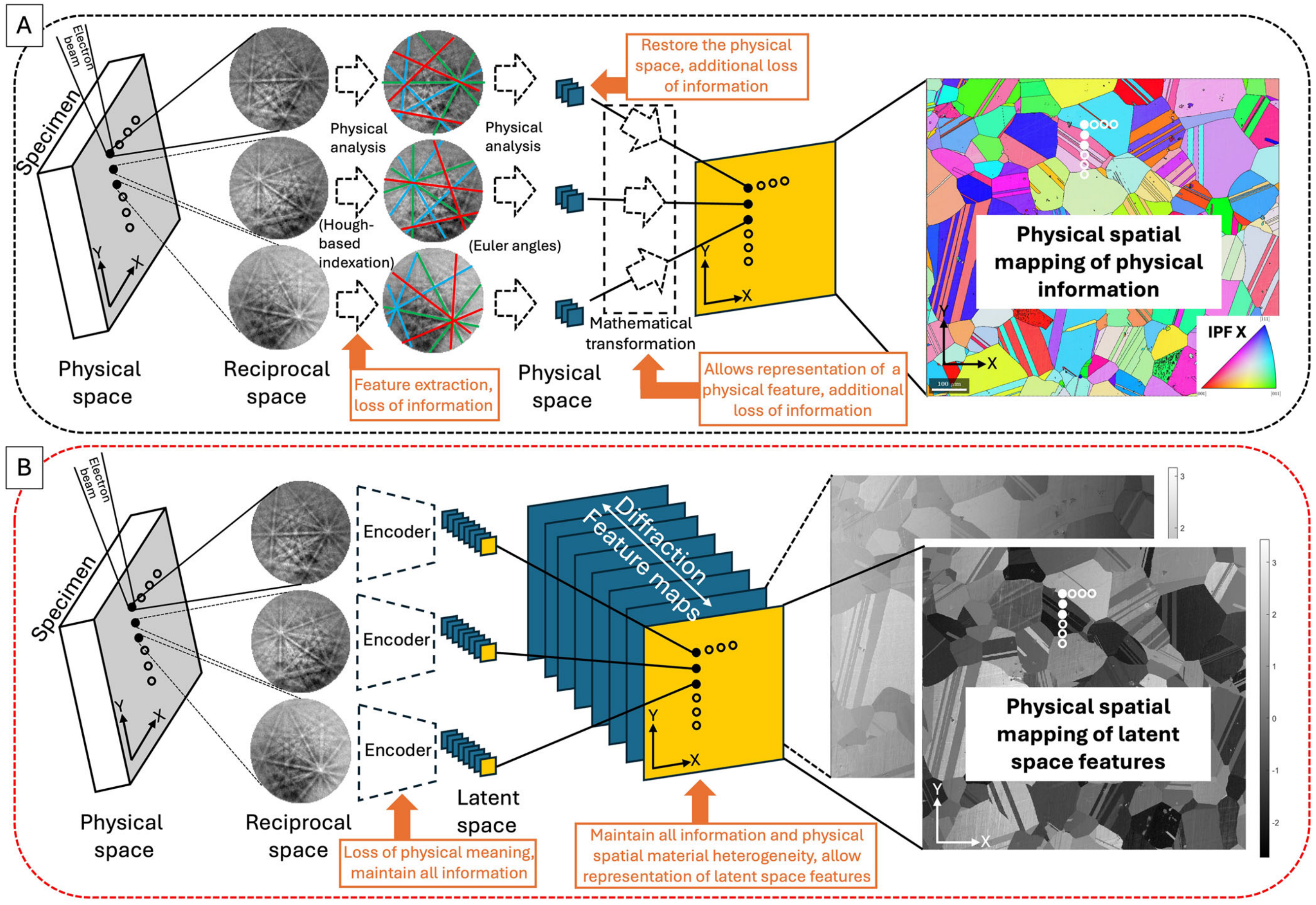

**Fig. 1 | Physics-based and data-driven processing of Kikuchi patterns for mapping microstructural heterogeneity. A** Physics-based approach to extract physical descriptors (Euler angle, here) from the reciprocal space and map microstructure heterogeneity through stereographic projection of physical directions (inverse pole figures). **B** The individual Kikuchi patterns are encoded to comprehensively map and highlight microstructural heterogeneity, point diffraction mapping is performed in physical space to capture the 2D reciprocal space at each point (Kikuchi patterns); each pattern is then encoded into a latent space, generating a set of diffraction latent space features that describe the patterns comprehensively; finally, these latent space features are spatially mapped back to the physical domain, creating a representation of microstructural heterogeneity within the latent space.

Different approaches then allow to encode (data-reduced representation) microstructure of metallic materials but often focus on crystallographic orientations or are generally limited to a few descriptors. In approaches such as graph networks or similar methods, the grain structure is generally simplified (grain size, and overall shape)[39–45]. Other critical microstructural information, such as dislocation density, dislocation heterogeneity or LAGB, are often absent or included as average values, missing their heterogeneity. Alternatively, two-point statistics can be used to describe the spatial distribution of phases (or features)[46–48], and spherical harmonics can be employed to describe the crystallographic orientation distribution[46,48,49]. However, these methods are not currently applied to hierarchical or multimodal data. More recently, variational autoencoders (VAE) opened new pathways for microstructure representations[50–55]. This way, each microstructure is defined by a low-dimensional latent space representation, which in principle allows for the conservation of all the information contained in the input microstructure. When considering experimental data, despite the capabilities and limitations of the previously mentioned methods, this first requires one or several characterization techniques able to capture the complexity of metal microstructures across different scales.

In this study, a novel data-driven approach is utilized to encode microstructure data, preserving the entirety of information contained within point diffraction data while capturing spatial microstructural heterogeneity. This approach, described in Fig. 1B involves, (i) encoding diffraction data into a latent space directly from Kikuchi patterns, avoiding limitations imposed by analyzing only specific physical descriptors; and (ii) mapping the encoded diffraction data (latent space features) across representative regions to comprehensively capture microstructural heterogeneity relevant to all embedded parameters. This methodology enables the direct identification of critical microstructural heterogeneities, providing an accurate and complete data-reduced representation of metal microstructure to enable the prediction of the mechanical properties of metallic materials.

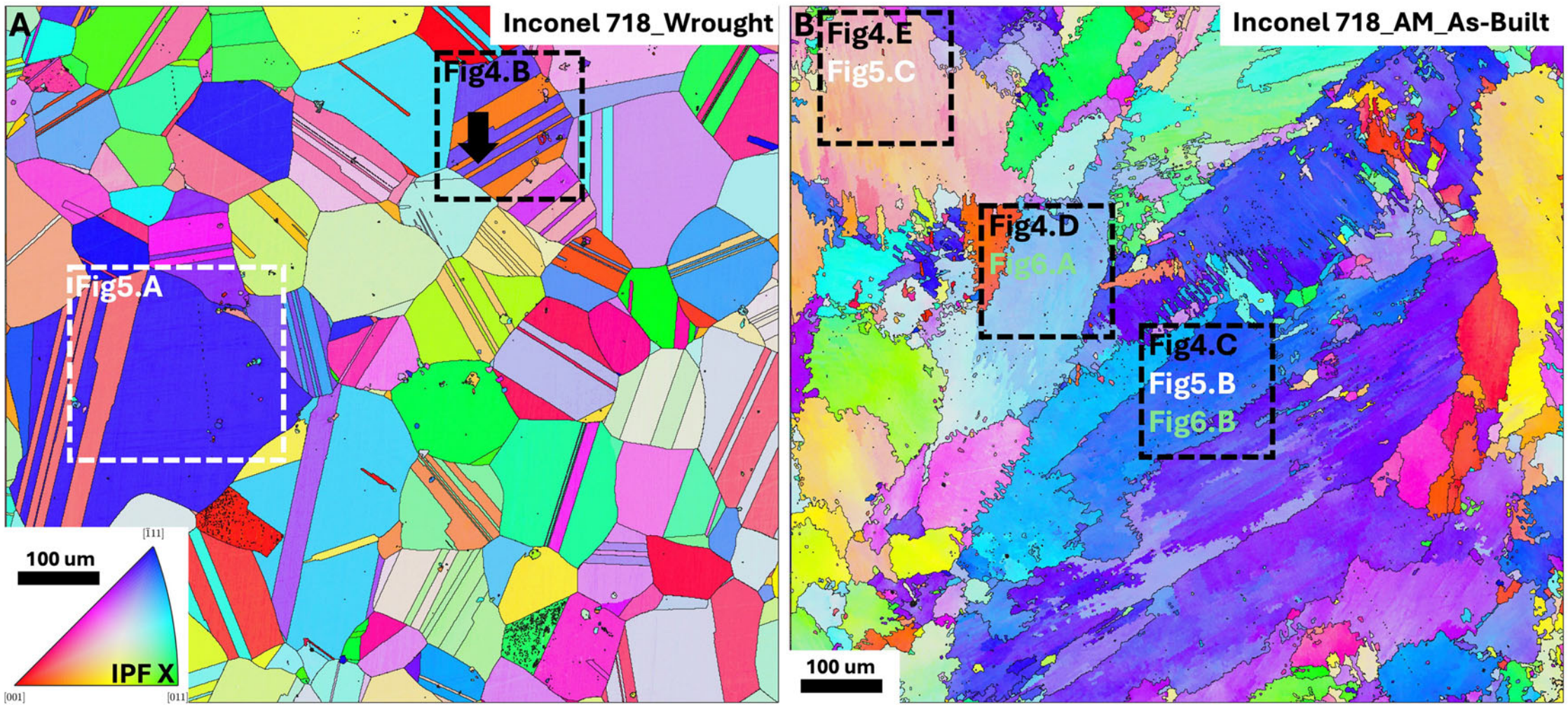

**Fig. 2 | Microstructure of investigated alloys.** Electron backscatter diffraction (EBSD) maps of the investigated Inconel 718 alloys. Inverse pole figure (IPF) maps along the X direction (horizontal) are presented for a **A** wrought and fully recrystallized 718 alloy, and a **B** additively manufactured as-built 718 alloy.

We investigate the use of VAE and contrastive learning to effectively encode Kikuchi patterns. The influence of hyperparameters is analyzed and discussed. The developed approach is applied to both a wrought and an AM fabricated superalloy, successfully identifying the key microstructural heterogeneities present in these materials. As expected, in wrought materials, the primary heterogeneities were associated with grain structure and crystallographic orientation. In contrast, in AM materials, our approach uncovered small-scale heterogeneities, such as variations in dislocation density (*i.e.*, cellular structures). Compared to conventional physics-based methods, the proposed approach demonstrates unprecedented sensitivity in detecting microstructural features. This enables the first comprehensive mapping of microstructural heterogeneities in AM metals. Finally, we discuss the application of this approach in predicting mechanical properties and guiding the design of new microstructures through the latent space.

## RESULTS

### Diffraction data

The dataset utilized in this study consists of conventional EBSD measurements taken over a large area (1 $mm^2$), following the procedure outlined in the methods section. The materials investigated include nickel-based superalloys processed via forging and AM, specifically a wrought, fully recrystallized Inconel 718 and an additively manufactured (AM) Inconel 718 in its as-built condition. Detailed descriptions of the investigated materials can be found in the methods section. The inverse pole figure (IPF) maps along the horizontal direction for the examined alloys are shown in Fig. 2. Significant discrepancies in microstructure morphology and heterogeneities between the wrought and AM materials are observed. The wrought material exhibits equiaxed crystallographic grains with uniform crystallographic orientation within each grain. In contrast, due to the unique thermal history and rapid solidification associated to the AM process, the AM Inconel 718 displays elongated grains, large lattice rotation gradients, and a high fraction of low angle grain boundaries. Additionally, a high density of dislocations[56], residual stresses[57] and potential chemical fluctuations are expected to be present[58–60]. During the EBSD measurements, diffraction patterns, known as Kikuchi patterns, and the position at which the patterns are taken, are recorded and stored as TIFF images with a resolution of 480 by 480 pixels. All patterns of a single EBSD map are stored as a UP2 file. Details on the acquisition and patterns themselves are provided in the methods section. These data have been made available in ref. 61.

As schematically illustrated in Fig. 1B, the proposed method begins by encoding the Kikuchi patterns, which requires the development and training of a specific machine learning approach. This process is detailed in the following section. Once the encoder is properly trained, it is applied to the Kikuchi patterns captured from both materials. All patterns are then encoded into a vector of latent feature values, which are subsequently spatially mapped onto the same grid from which the associated patterns were collected.

### Encoding of diffraction data

A machine learning architecture has been developed and trained to encode experimental Kikuchi patterns into a low-dimensional latent space representation. For that purpose, we followed and tested two distinct approaches, (a) a variational autoencoder (VAE)[62] and (b) a modified VAE including a contrastive learning approach (SimCLR) initially developed for classification purposes[63]; the complete structure of the architecture designed for this application is given in Fig. 3. Two different convolutional neural networks (CNN) are used, an *Encoder* used to convert the Kikuchi patterns to their latent space representations and a *Decoder* to restore the Kikuchi patterns from the low-dimensional representations. Different numbers of latent space dimensions have been investigated, ranging from 16 to 256 dimensions and the associated numbers of kernels are detailed in Table 2 in the methods section.

When training machine learning models, data augmentation is an efficient way to increase the amount of data seen by the architecture. However, not all types of augmentation commonly used in the literature are suitable for augmenting Kikuchi patterns. Some augmentations used in the original SimCLR study are not suitable for Kikuchi patterns augmentation, including crop, resize, cutout, rotation or Gaussian blur as mentioned in Ref. 64. For instance, rotations would affect the crystallographic orientations, blur would affect the 'sharpness' and therefore features related to the dislocation density.

Figure 3A illustrates the training principles of the VAE. The loss function consists in a pixel-to-pixel $L_2$ loss calculated between the original pattern and its reconstructed counterpart (once encoded and decoded). The region surrounding the patterns (artificially black region with no signal) has been masked out and therefore does not contribute to the evaluation of the loss. A Kullback-Leibler divergence term is also added to ensure that the learned distributions converge toward a standard normal distribution[65]. The dataset used for training consists of 96,000 randomly selected Kikuchi patterns from all investigated materials. No augmentations have been used to train this conventional VAE architecture.

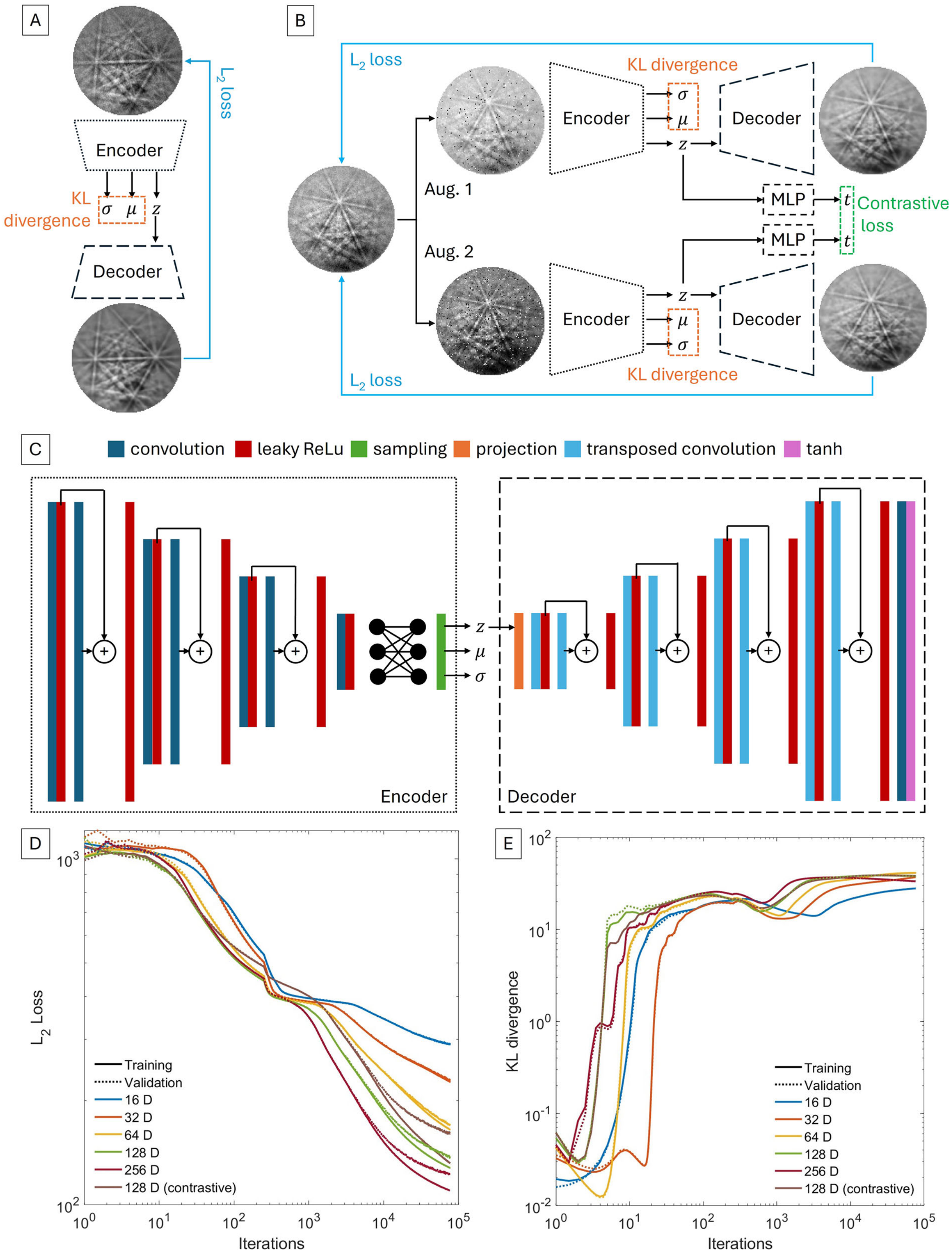

**Fig. 3 | Architecture and training for Kikuchi pattern encoding.** **A** Training principle of the variational autoencoder and associated loss functions used in this study. **B** Parallel variational autoencoder processing augmented versions of the same set of Kikuchi patterns and integrating a contrastive loss[63,66]. **C** Detailed architectures of the *Encoder* and *Decoder* convolutional neural networks used in both the A and B approach. **D** $L_2$ loss and **E** KL divergence evolutions throughout training for the conventional variational autoencoder and parallel configuration.

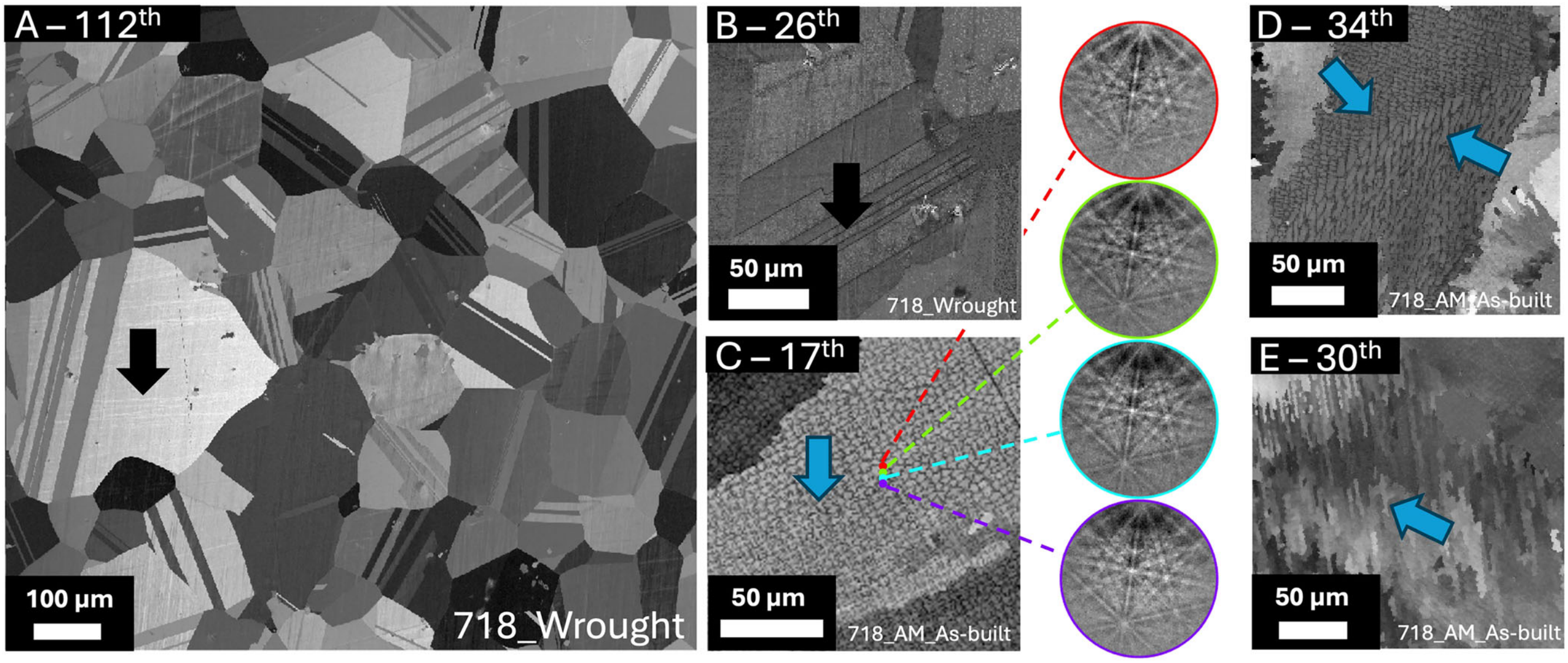

**Fig. 4 | Latent space feature maps for microstructure heterogeneity identification.** **A** The 112th latent space feature map for the wrought 718 material using the 128-dimension latent space architecture and VAE training. **B** A reduced region of the 26th latent space feature map for the wrought 718 material. **C, D** AM Dislocation cells revealed by the encoding of Kikuchi patterns using the 256-dimension latent space architecture in the AM as-built material. The 17 and 34th latent space features are represented, respectively. **C** Series of Kikuchi patterns associated along a path intersecting a dislocation cell. **E** 30th latent space feature map of a reduced region of interest for the AM as-built material showing rotation domains within a grain.

Concurrently and to evaluate a different training procedure and with the goal of being more robust against acquisition noise, we integrate an additional contrastive loss[63,66] within the VAE approach, as shown in Fig. 3B. In the original SimCLR approach[63], developed for classification purposes, features are extracted by a ResNet architecture[67] from two augmented versions of the same image and projected using a multilayer perceptron (MLP). In the present study, the training principle involves three different augmentations related to possible acquisition noise, salt and pepper noise, Gaussian noise and Gamma alteration. Contrary to the augmentation paths proposed in SimCLR[63], the previously mentioned augmentations have been superimposed on the same pattern. In contrast to common VAE data augmentation, these augmentations serve a different purpose as the network is trained to reconstruct the input image (i.e., without augmentation). For this training, we used a pixel-to-pixel $L_2$ loss calculated between the original pattern (before augmentations) and the reconstructed versions. A Kullback-Leibler divergence is also evaluated for each encoded set of augmented Kikuchi patterns. Finally, a contrastive loss term is added based on these projected representations[66]. The weight of this contrastive loss has been multiplied by 20 to obtain values comparable with the KL divergence. In our approach, the low-dimensional representations are not used for classification but need to be suitable for pattern reconstruction. This additional loss term helps the architecture to produce a robust representation within the latent space despite the various noise that may be added through the augmentations to the initial image. The *Encoder* and *Decoder* share the same weights between the two augmentation paths.

The evolution of the $L_2$ loss and KL divergence during training of the conventional VAE architecture are shown in Fig. 3D, E, respectively. The metrics corresponding to the training sets are depicted as solid lines while the ones associated to the validation sets are plotted as dashed lines. All the metrics have been smoothed over a 500-iteration window for visualization purposes. As expected, increasing the number of dimensions allows to significantly reduce the pixel-to-pixel loss as the latent space is able to store more information. For most of the number of dimensions considered, no difference is observed between the training and validation sets. The training was discontinued after 250 epochs (76,500 iterations) for all VAE architectures, as the validation $L_2$ loss showed early signs of over-fitting (emphasized by the log-scale of the plots). Concerning the parallel VAE configuration, the presented metrics evolution corresponds to the metrics average between the two branches and augmentation paths. Its loss indicates an intermediate evolution between conventional 64- and 128-dimension VAEs. However, it is worth noting that the loss is established with the original pattern, which could explain the difference in loss. Compared to the conventional 128-dimension VAE, the parallel configuration exhibits earlier signs of over-fitting. Regarding the KL divergence evolution (Fig. 3E), no significant difference is observed between the training and validation sets and all converge towards a similar value at 250 epochs including the parallel VAE configuration.

## Latent space features mapping

After training the *Encoder* using either the conventional VAE or our SimCLR approach, we utilize it to encode all collected Kikuchi patterns from both investigated materials. This process provides an efficient solution to build a low-dimensional representation (a vector of 16, 32, 64, 128, or 256 latent space features) while mitigating the loss of information. These features are then spatially mapped according to the physical grid used during the collection of the Kikuchi patterns. As a result, for both materials, we generate 16, 32, 64, 128, or 256 maps that can be represented and visualized.

The map of the 112th latent space feature from the 128-dimensional latent space architecture is shown in Fig. 4A for the wrought 718 material. This feature map reveals the grain structure, which closely corresponds to the grain morphology observed in the IPF map in Fig. 2A. Notably, within the reduced region of interest depicted in Fig. 4B for the wrought 718 material, the 26th latent space feature shows a very thin annealing twin (indicated by the black arrow) that is not discernible in the IPF map in Fig. 2A. In Fig. 4C, D, maps of the 17 and 34th latent space features are presented for two reduced regions of interest in the 718 AM as-built material. The regions of interest, as indicated by dashed-line boxes, are shown in the IPF maps in Fig. 2B. Interestingly, while heterogeneity at the small scale is not directly observable in the IPF map, encoding the diffraction data into the latent space reveals small-scale heterogeneities (blue arrows in Fig. 4C, D) associated with the AM dislocation cellular structure[11]. The sensitivity of the encoder is such that, within the grain shown in Fig. 4D, variations in both the size and geometric orientation of the AM cells are clearly visible. Figure 4C offers a collection of original Kikuchi patterns chosen along a path intersecting a dislocation cell wall. These patterns do not visibly show differences but theoretically contain the information about dislocation density[11] which is effectively revealed by the encoding of these patterns. Finally, Fig. 4E demonstrates the encoder's sensitivity in detecting small-scale rotation domains associated with the AM solidification process[11] within an AM grain.

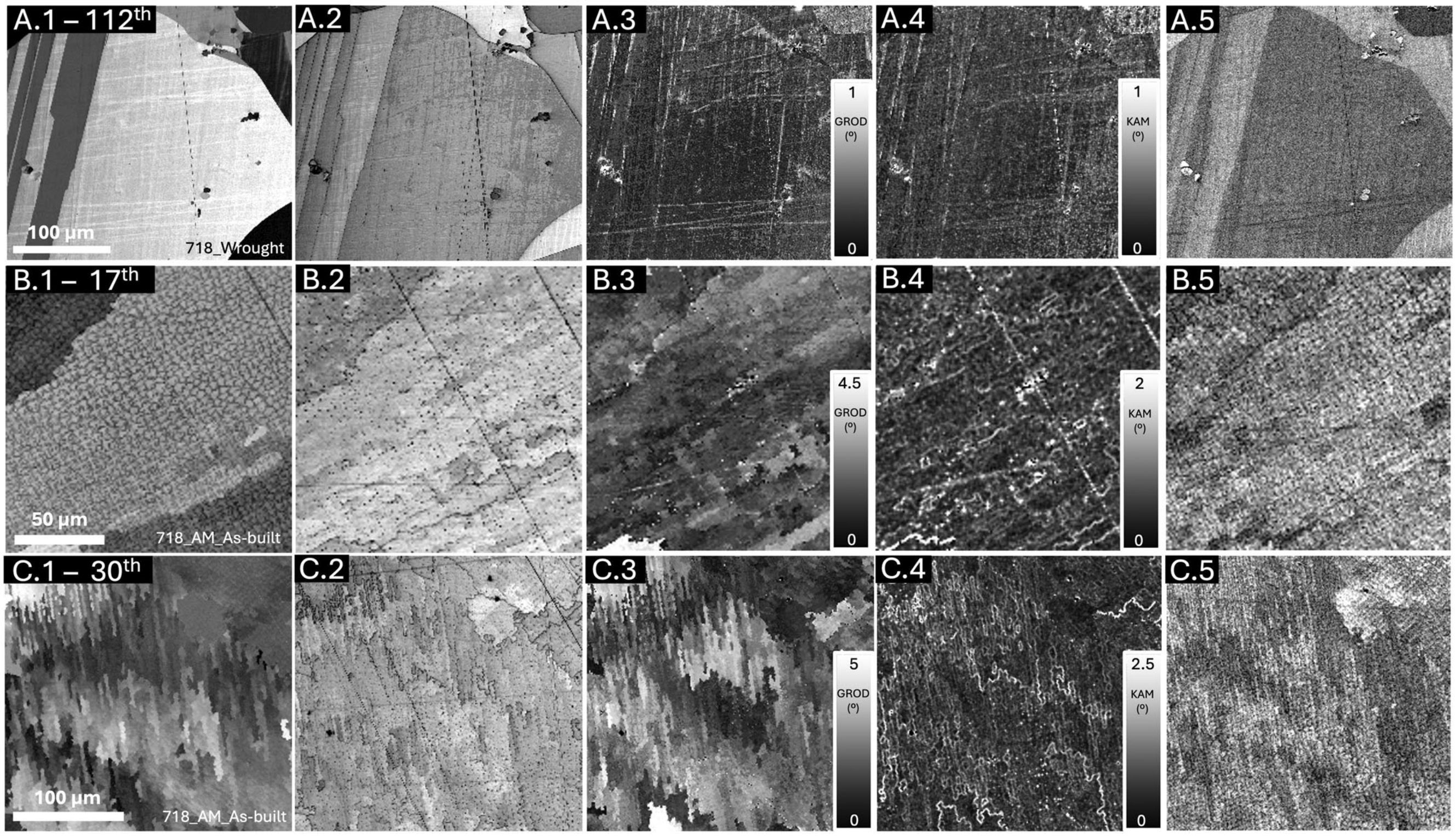

**Fig. 5 | Comparison of latent space features maps and conventional EBSD. A.1** The 112th latent space feature map of a region of interest for the wrought 718 material using the 128-dimension latent space architecture and VAE training. **(A.2-A.5)** Associated maps from Kikuchi pattern feature extraction and mathematical transformations traditionally employed in EBSD analysis, **(.2)** image quality; **(.3)** grain reference orientation deviation; **(.4)** kernel average misorientation; **(.5)** sharpness. **B.1,C.1** The 17 and 30th latent space feature maps of two regions of interest for the AM as-built 718 material. **B.2–B.5,C.2–C.5)** Associated maps from Kikuchi pattern feature extraction and mathematical transformations traditionally employed in EBSD analysis, **(.2)** image quality; **(.3)** grain reference orientation deviation; **(.4)** kernel average misorientation; **(.5)** sharpness.

Several maps derived from conventional physical analyses of Kikuchi patterns are presented for the wrought 718 material (A) and the AM as-built material (B, C) in Fig. 5(A.2-A.5, B.2-B.5 and C.2-C.5). These maps originate from physical Kikuchi pattern feature extraction and mathematical transformations traditionally employed in EBSD analysis, including, (i) image quality (IQ), representing the average intensity of the Kikuchi pattern (A.2, B.2, C.2); (ii) grain reference orientation deviation (GROD), which quantifies point misorientation within a grain relative to its average grain orientation (A.3, B.3, C.3); (iii) kernel average misorientation (KAM), representing point misorientation as a function of neighboring points (A.4, B.4, C.4), and (iv) sharpness[11,68], indicating the diffuseness of Kikuchi bands (A.5, B.5, C.5). These maps are compared to a latent space feature map displayed in Fig. 5(A.1, B.1, C.1) for the same regions of interest. The regions of interest shown in Fig. 5 correspond to the dashed boxes in Fig. 2.

The sensitivity of the 112th latent feature is associated with the crystallographic orientation, as evidenced by the grain structure revealed in Fig. 5A.1. However, heterogeneity (indicated by brighter lines within the bright grain) are also observed, as shown in Fig. 5A.1. This heterogeneity corresponds to residual plasticity introduced during sample preparation, due to residual dislocations from surface scratches. Notably, such heterogeneities are not clearly discernible in the physically based analyses maps shown in Fig. 5A.2–A.5. Similarly, the AM cellular structure is prominently visible in the 17th latent feature map in Fig. 5B.1, while it is not captured in the traditional analysis maps in Fig. 5B.2–B.5. Even the sharpness analysis, a specialized physical analysis designed to extract dislocation cells, as shown in Fig. 5B.5, fails to clearly highlight the cell structure. In Fig. 5C.1, the 30th latent feature map captures small-scale AM orientation domains, which are also observed (with varying contrast) in the corresponding GROD map shown in Fig. 5C.3.

Additional latent space feature maps for two regions of interest in the AM as-built 718 material are presented in Fig. 6A.1, A.2, B.1, B.2. While the 34 and 17th latent space features demonstrate high sensitivity to the AM dislocation cellular structure, the 13th latent space feature is notably sensitive to crystallographic orientation. This sensitivity is evidenced by the similar variations observed in the corresponding GROD maps shown in Fig. 6A.3, B.3.

The previous maps were generated from EBSD measurements conducted with a step size of 1 µm, whereas the measurements shown in Fig. 7 were performed with a smaller step size of 0.1 µm. The region of interest corresponds to the area highlighted by the dashed white box in Fig. 6A.1. The 56 and 158th latent space features, extracted using a 256-dimensional latent space architecture and VAE training, reveal the AM dislocation cellular structure with unprecedented resolution. Compared to the sharpness analysis shown in Fig. 7B, D, the 56th latent space feature map reveals detailed structures within the cell walls. The 128th latent space feature map in Fig. 7A.2, C.2 demonstrates sensitivity to both crystallographic misorientation between cells and the AM dislocation cellular structure. Interestingly, contrasts within the cell are observed. Moreover, small-scale particles (Ni-, Nb-rich $Ni_2Nb$ Laves phase, nitrides and oxides; visible as bright white spots) are also observed on the 158th latent space feature map, corresponding to particles forming along the cell walls in 718[69]. These particles and the contrast variations within the cell or along the cell wall are not visible in the GROD, KAM, IQ (not reported here), or sharpness maps derived from conventional EBSD analysis.

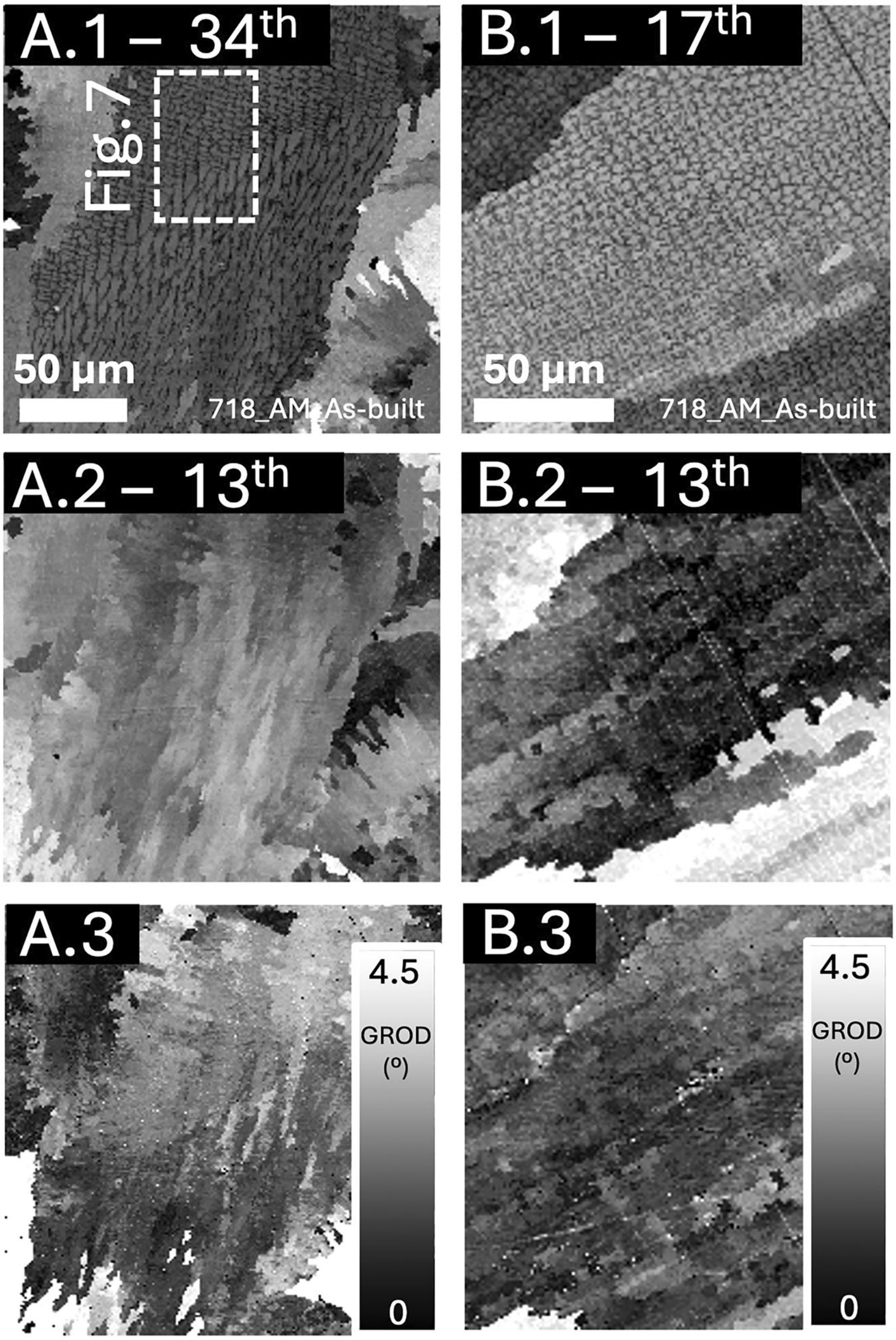

**Fig. 6 | Comparison of latent space features maps and conventional EBSD. A, B** Latent space feature maps (**.1-.2**) and GROD maps (**.3**) for two regions of interest in the AM as-built materials. These regions are highlighted with dashed boxes in Fig. 2. (**A.1-A.2, B.1-B.2**) Display of two distinct latent space feature representations for the two regions of interest.

### Latent space microstructure design, latent space and physical features disentanglement

The latent space associated to the low-dimensional representations depends on both the *Encoder* and *Decoder* architectures but also the loss functions used during training. First, we trained conventional VAEs[62] with two loss terms (pixel-to-pixel $L_2$ loss and KL divergence) and various number of latent space dimensions ranging from 16 to 256. With limited numbers of latent space dimensions, the architecture first focuses on the reconstruction of main features such as the background and the bands as illustrated in Fig. 8B.1. With more dimensions, the latent space is able to store more information and provides better reconstructions as shown in Fig. 8B.2–B.5 for the same diffraction patterns. In the first row of Fig. 8A.1–A.5 is given the difference with the preprocessed pattern. With sufficient dimensions, the only difference that remains is not associated to the bands but to the random noise of the experimental pattern. This evolution of reconstruction quality confirms the observations relative to the $L_2$ loss (Fig. 3D), which is significantly reduced when the number of dimensions of the latent space is increased. Additionally, Fig. 8C illustrates the difference between reconstruction while adding more dimensions to the latent space representation. Going from 16 to 32 dimensions improves band reconstruction in the center of the pattern. Then, going from 32 to 64 dimensions improves the bands near the outside of the pattern. Advancing to 128 and 256 dimensions consists of smaller improvements about three times less intense compared to the 16–32 change and localized on bands but also in between bands.

To disentangle the various physical features within the dimensions of the latent space, one approach is to encode the data into a latent space with more dimensions. The feature maps most visually sensitive to the cell structure were extracted using a VAE architecture with latent space dimensions of 32, 64, 128, and 256, as shown in Fig. 9A–D, respectively. It is observed that increasing the dimensionality of the latent space improved the visualization of the cell structure.

At the end of the training of the VAE architecture (128 dimensions), reconstructions considered as best and worst (in terms of $L_2$ loss) are given in Fig. 10A.1, A.2, respectively. Both qualities of reconstruction show patterns of acceptable quality for crystallographic orientation determination. Additionally, these VAE architectures can also be used for generating a new image, here a pattern, from latent space values[62]. For this purpose, a random latent space representation is generated from unit normal distributions and processed through the *Decoder*. Accordingly, Fig. 10A.3 illustrates randomly generated patterns with realistic background. However, the decoder generates bands that are not always straight and unrealistic poles (intersections of bands).

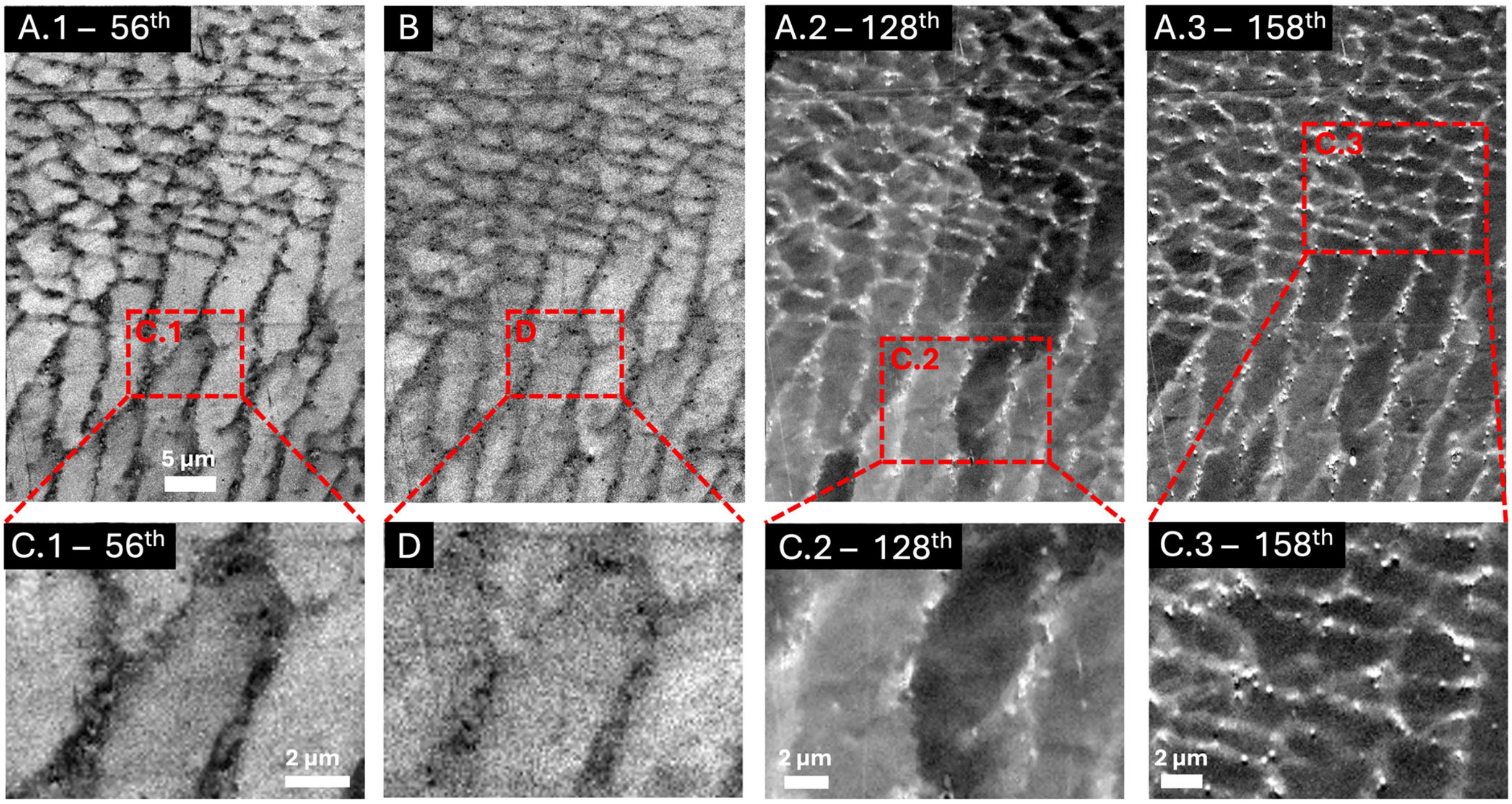

**Fig. 7 | High-resolution latent space feature maps. A.1-A.3** High-resolution representation of the 56, 128, and 158th latent space feature maps for the reduced region of interest in Fig. 6A.1. **B** The associated sharpness map. **C.1–C.3** Reduced regions of interest of the latent space feature maps. **D** Reduced regions of interest for the sharpness map.

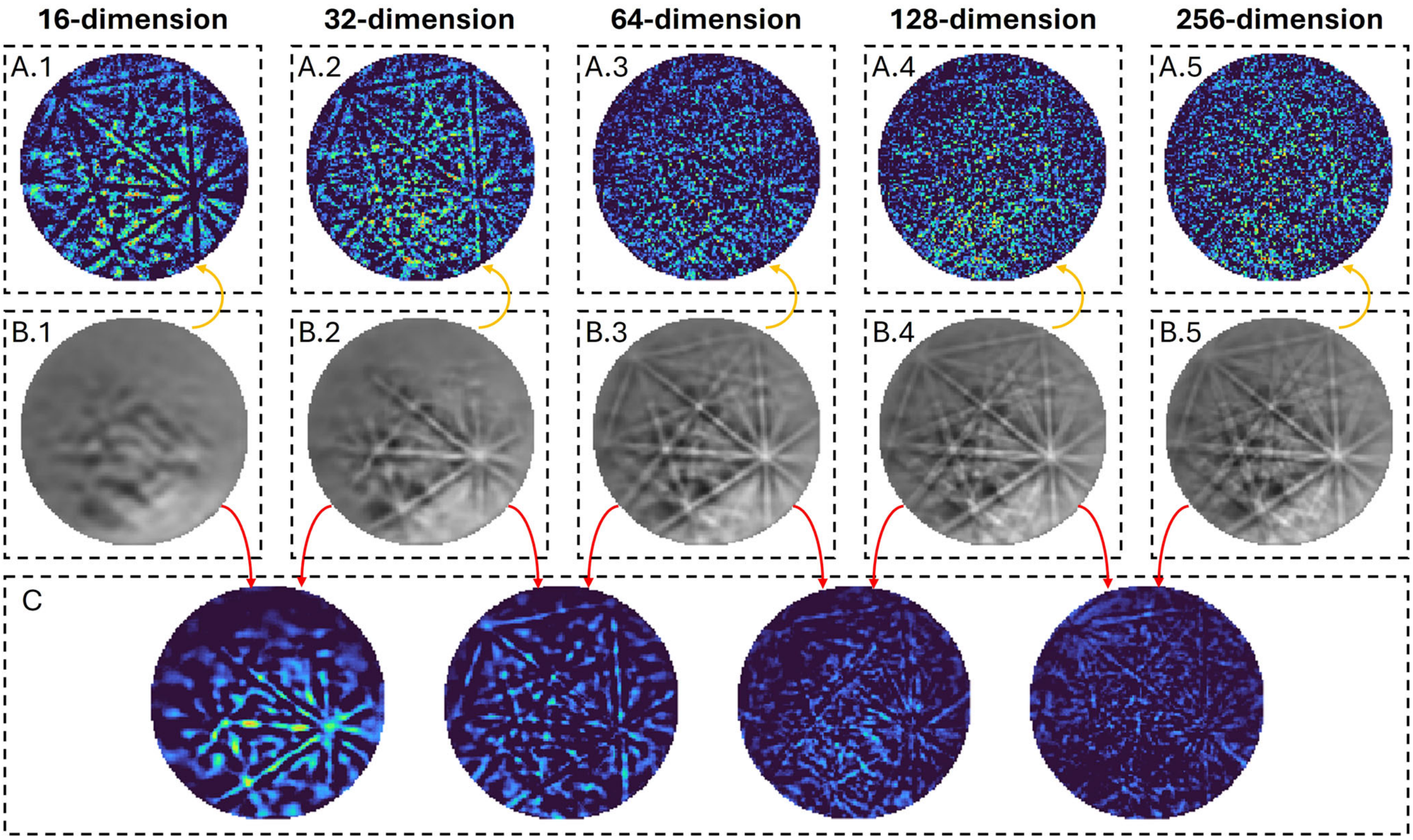

**Fig. 8 | Impact of latent space dimensionality. A** Pixel-to-pixel differences with a given original pattern and **B** corresponding patterns once encoded/decoded by the VAE architectures, (**.1**) 16-dimension, (**.2**) 32-dimension, (**.3**) 64-dimension, (**.4**) 128-dimension and (**.5**) 256-dimension. **C** From left to right, the pixel-to-pixel differences between the encoded/decoded patterns from a 32-dimension to a 16-dimension; a 64-dimension to a 32-dimension; a 128-dimension to a 64-dimension; and a 256-dimension to a 128-dimension. All differences are shown using the same scale.

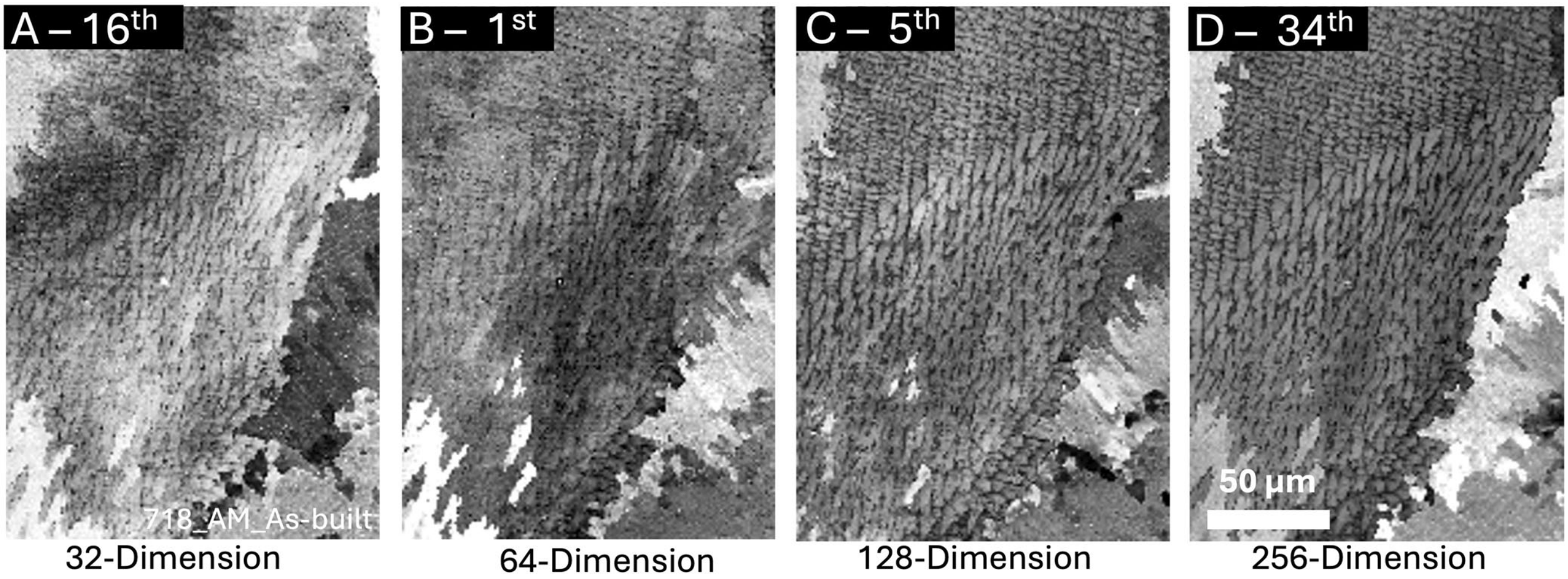

**Fig. 9 | Impact of latent space dimensionality.** The latent space features that reveal the best the cellular structure within the the as-built 718 material using the **A** 32-dimension, **B** 64-dimension, **C** 128-dimension, and **D** 256-dimension latent space architecture and VAE training.

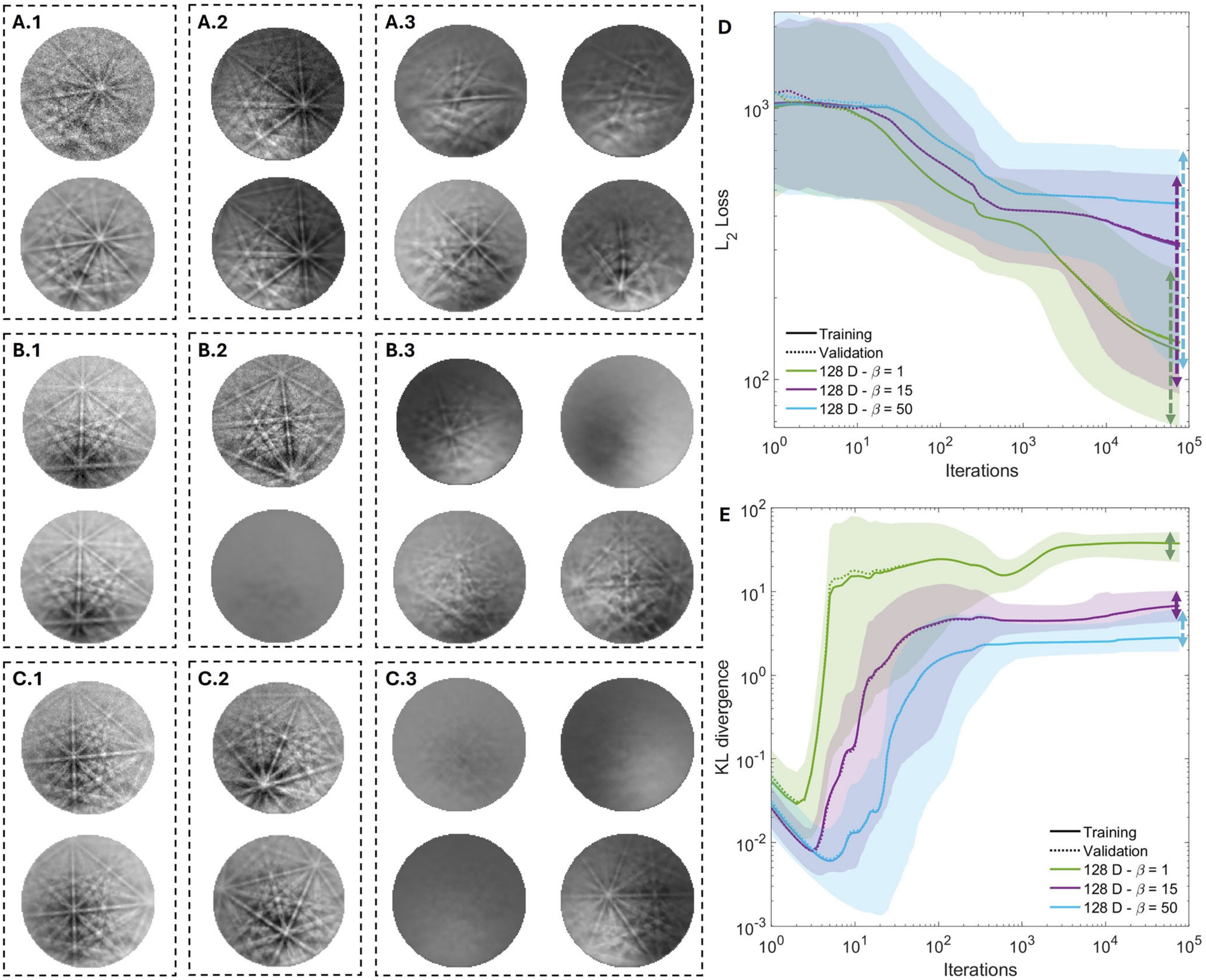

**Fig. 10 | Effect of $\beta$ regularization.** Kikuchi patterns associated to **A** conventional VAE ($\beta = 1$), **B** $\beta$-VAE ($\beta = 15$) and **C** $\beta$-VAE ($\beta = 50$). Quality of reconstruction at the end of the training of these architectures (original pattern on top), **(.1)** best reconstruction and **(.2)** worst reconstruction. **(.3)** Associated randomly generated Kikuchi patterns from unit normal distributions. Associated **D** $L_2$ loss and **E** KL divergence evolutions throughout training for the different regularization parameter $\beta$.

**Table 1 | Performances of the various latent space dimensions and different loss functions**

| Loss type | Number of dimensions | Indexed patterns | Confidence index ratio 10% / 50% / 90% | Image quality ratio 10% / 50% / 90% | Misorientations (°) 10% / 50% / 90% | Sharpness ratio 10% / 50% / 90% |
|---|---|---|---|---|---|---|
| VAE | 16 | 95.3% | 0.00/0.06/0.42 | 0.27/ 0.64 / 0.87 | 0.01 / 0.02 / 0.88 | 0.13 / 0.16 / 0.18 |
| | 32 | 98.8% | 0.03 / 0.53 / 1.10 | 0.54 / 0.87 / 0.99 | 0.00 / 0.01 / 0.07 | 0.16 / 0.23 / 0.27 |
| | 64 | 99.9% | 0.50 / 1.00 / 1.78 | 0.80 / 0.95 / 1.03 | 0.00 / 0.01 / 0.02 | 0.23 / 0.27 / 0.32 |
| | 128 | 100% | 0.55 / 1.00 / 1.79 | 0.88 / 0.97 / 1.04 | 0.00 / 0.01 / 0.01 | 0.28 / 0.34 / 0.39 |
| | 256 | 100% | 0.56 / 1.00 / 1.78 | 0.88 / 0.96 / 1.02 | 0.00 / 0.01 / 0.01 | 0.32 / 0.38 / 0.44 |
| Contrastive | 128 | 100% | 0.56 / 1.00 / 1.79 | 0.84 / 0.95 / 1.05 | 0.00 / 0.01 / 0.01 | 0.27 / 0.33 / 0.38 |
| $\beta$-VAE | 128 ($\beta$ = 15) | 76.2% | 0.12 / 0.94 / 1.61 | 0.35 / 0.88 / 1.04 | 0.00 / 0.01 / 0.41 | 0.11 / 0.26 / 0.38 |
| | 128 ($\beta$ = 50) | 39.2% | 0.00 / 0.35 / 1.20 | 0.08 / 0.20 / 0.92 | 0.00 / 0.28 / 0.89 | 0.05 / 0.09 / 0.29 |

As another approach to enable disentangling of physical features, we used a $\beta$-VAE, an alternative to conventional VAE, developed by Higgins et al. for dimensions disentanglement[70]. Under these conditions, modifying only one coordinate within the low-dimensional representation should only affect one characteristic in the reconstructed image[70]. A regularization parameter $\beta$ is used to increase the weight of the Kullback-Leibler divergence within the loss function. Using the same training parameters, two $\beta$-VAE architectures were trained with a 128-dimension latent space and using $\beta$ values of 15 and 50.

Once compared to the results associated to the conventional VAE for which $\beta = 1$ (Fig. 10A), increasing $\beta$ produces a comparable quality of reconstruction for the best reconstructions (see Fig. 10B.1, C.1), however this significantly affects the quality of the worst reconstructions (see Fig. 10B.2, C.2. The loss associated with the worst reconstruction remains constant from halfway through training, indicating that longer training would not improve the reconstruction quality of these patterns. At the end of the training, the worst reconstructions only show a background without any recognizable pattern in Fig. 10B.2 associated to $\beta$ values of 15. For a $\beta$ value of 50, two scenarios can occur with the worst reconstruction, either a background with no recognizable pattern or an erroneous reconstruction as shown in Fig. 10C.2. As with conventional VAE, the Kikuchi patterns were decoded from a random latent space representation. For a $\beta$ value of 15, randomly generated patterns can be radically different, from a realistic Kikuchi pattern (top left in Fig. 10B.2) to a pattern with no bands (top right in Fig. 10B.3). The remaining patterns show straight but blurry bands. Concerning the highest considered $\beta$ value of 50, most of the randomly generated patterns do not show any band (see Fig. 10C.3).

In addition to the reconstructed patterns, the $L_2$ loss and KL divergence associated to the training of these architectures are given in Fig. 10D, E, respectively. The average metrics are indicated by a solid line for the training set and by a dashed line for the validation set. The range of these metrics are plotted by shaded regions extending from the 1% to the 99% patterns (in terms of $L_2$ loss) and illustrated by arrows. All $L_2$ loss metrics show a similar evolution. However, the average value at 250 epochs increases significantly with the regularization parameter. While the lowest $L_2$ loss (bottom bound of the shaded region) is only moderately affected by a $\beta$ increase, the poor reconstructions (upper bound of the shaded region) demonstrate a substantial increase in $L_2$ loss. This increase manifests as patterns without bands or erroneous reconstructions, as illustrated in Fig. 10B.2, C.2. Regarding the KL divergence evolution (Fig. 10E), Although the evolution is comparable, increasing $\beta$ leads to a delay in the KL divergence evolution. At 250 epochs, the KL divergence is significantly reduced, providing efficient leverage for reducing the total loss.

Using all VAEs trained in this study, patterns have been processed through the complete architectures, reassembled into UP2 files using KikuchiPy[71] and indexed using EDAX OIM 8 software without any additional background correction or histogram modification. The Hough transform has been computed using the same parameters, $\theta$ step of 1° and $\rho$ fraction of 85%. The bands found in the Kikuchi patterns are identified as peaks within the Hough transform with a minimum magnitude of 5 and minimum distance of 10. EBSD maps were finally compared to the original ones (preprocessing detailed in the methods section) within MATLAB using the MTEX toolbox[72]. The performances are given in Table 1 in terms of misorientation, IQ ratio, CI ratio and sharpness ratio. For that, the characteristics identified from encoded/decoded patterns have been divided (or compared) by the ones associated to the preprocessed pattern and are reported in terms of 10, 50 (median) and 90$^{th}$ quantiles of the associated distributions. The corresponding EBSD and IQ maps are shown in Fig. 11, and can be compared to the preprocessed results given in Fig. 11A.1, B.1.

As expected, the 16-dimension latent space gives the worst reconstruction among the conventional VAEs, and consequently strongly affects all the performances. Despite the obviously poor quality of the reconstructed pattern as shown in Fig. 8B.1, 95.3% of the diffraction patterns were indexed with a limited misorientation of 0.88° for the 90th quantile. Apart from the grains shown in blue, most of the grains in Fig. 11b.1 show non-indexed patterns. From 64 dimensions, 99.9% of the patterns were indexed with orientations very similar to the preprocessed patterns, 90th-quantile misorientation is only 0.02°. The CI associated with these patterns is greatly enhanced, and the IQ is slightly improved. The best performances are obtained with the 256-dimension VAE, for which the corresponding EBSD and IQ maps are shown in Fig. 11C.1, C.2, respectively. Although very similar, the IQ appears to be slightly lower than its original counterpart. When compared to conventional 128-dimension VAE, the contrastive variant provides the same performances for the given characteristics. We also investigated an alternative to conventional VAE with $\beta$-VAE for dimension disentanglement. While the quality of the best reconstructions appears to be comparable to conventional VAE at the end of training, some patterns show poor reconstruction with no bands (Fig. 10B.2, C.2). Similar conclusions can be drawn from the performances, some patterns seem to be improved compared to the preprocessed pattern (high 90th-quantile CI), however only 76.2 and 39.2% of the patterns were indexed with $\beta = 15$, $\beta =;50$, respectively. The larger the regularization parameter $\beta$, the larger the latent space domain leading to a poor reconstruction. Such results can be visualized in Fig. 11D.1, E.1, where some specific orientations are not indexed at all, while the diffraction patterns associated with the grains shown in blue are well reconstructed and indexed, most likely due to the significant contribution of these orientations in the training data. Furthermore, increasing $\beta$ can also lead to an erroneous reconstruction as shown in Fig. 10C.2 where the reconstructed pattern is of sufficient quality to identify the crystallographic orientation, but does not match the original pattern. This can be specifically observed within Fig. 11E.1 where some grains present a noisy but globally different crystallographic orientation when compared to Fig. 11A.1. Ultimately, the training of $\beta$-VAE leads to a trade-off between the loss associated to the quality of reconstruction and the loss forcing the distributions to converge towards standard normal distribution. For a large $\beta$ value, mitigating the loss associated with the KL divergence becomes more valuable to the network than improving the reconstruction itself.

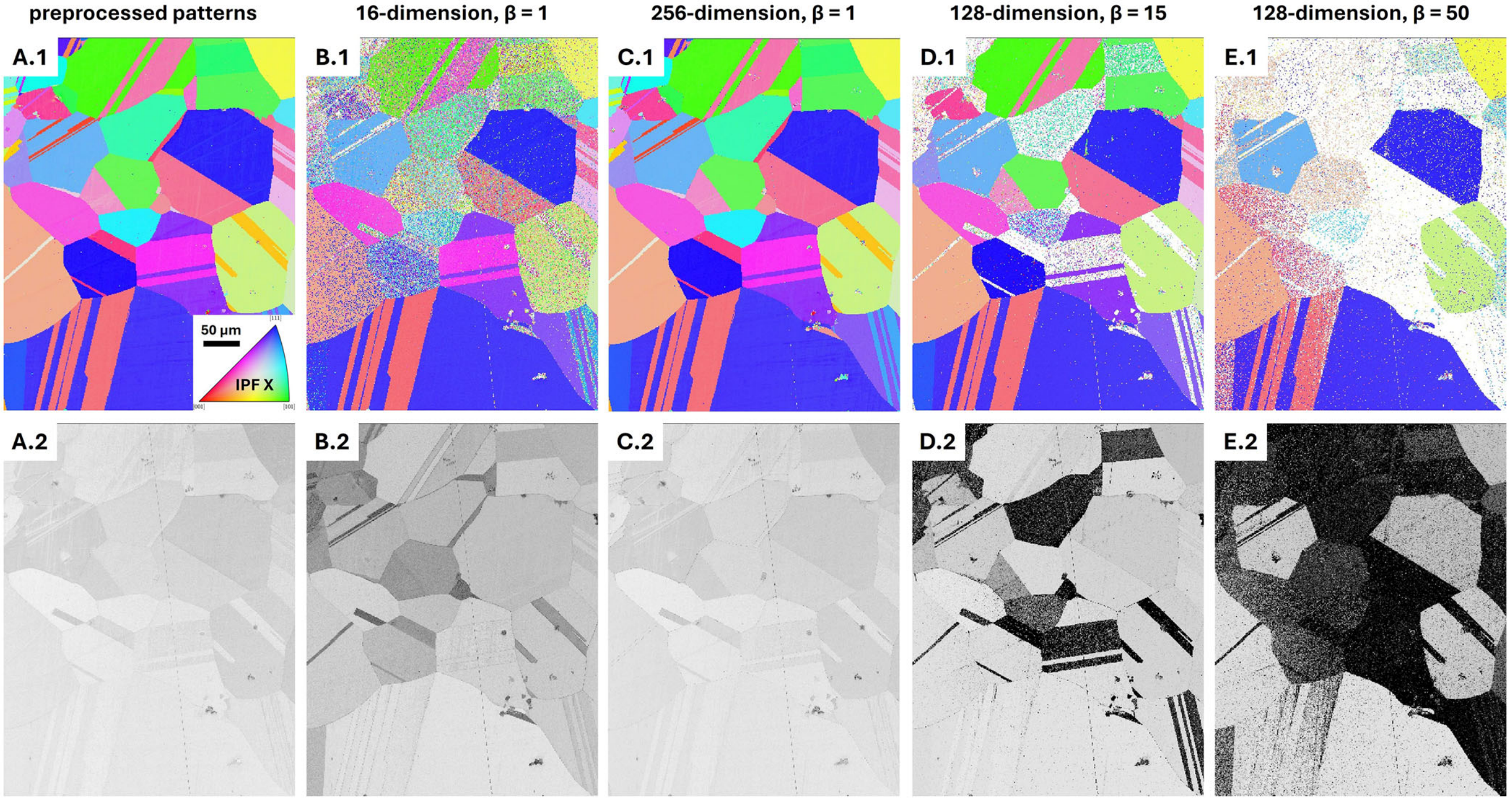

**Fig. 11 | Dimensionality and regularization effects on pattern indexing.** Results associated to **A** preprocessed patterns and **B–E** encoded/decoded patterns from different latent space, **B** 16-dimension with $\beta = 1$ (default), **C** 256-dimension with $\beta = 1$ (default), **D** 128-dimension with $\beta = 15$ and **E** 128-dimension with $\beta = 50$. (**.1**) Orientations maps and (**.2**) image quality.

## DISCUSSION

CNNs have been successfully applied to Kikuchi patterns for tasks such as identifying symmetry and crystallographic orientation[73–76]. They have facilitated the physical processing of Kikuchi patterns to determine crystal symmetry[73,74], advancements in indexing[64], phase identification[75,77], deformation state analysis[78], denoising[79], and handling high-noise diffraction patterns[76]. However, to the authors' best knowledge, all machine learning-based analyses of metal diffraction to date have been focused on aiding the extraction of known physical features (i.e., crystallographic orientation, dislocation density, and phase information). In the present approach, we encode diffraction patterns to capture all theoretical microstructural information contained within the Kikuchi patterns, rather than limiting the analysis to predefined physical features. A second key innovation is to map the low-dimensional representation to rapidly identify microstructural heterogeneities. Traditionally, this has been achieved in materials science through physical processing of EBSD measurements[80], or via more time consuming experimental techniques, such as electron channeling contrast imaging (ECCI)[81,82] and energy-dispersive X-ray spectroscopy (EDS)[83]. For instance, ECCI[84,85] has been effective in identifying small scale features such as cellular structures, whereas EBSD is primarily used for phase or crystallographic orientation identification at a larger scale and has not been extensively utilized for identifying small microstructure features. Similarly, EDS is commonly employed for chemical segregation analysis, despite the fact that chemical fluctuation information is inherently present within Kikuchi diffraction patterns (structure factor and lattice parameter effect). We usually rely on complex multimodal or correlative[86] measurement limiting large field of view measurement and therefore preventing the comprehensive characterization of the multi-scale heterogeneity of metallic materials. More importantly these datasets are too large and complex to be encoded or require too much time to be collected for effective and rapid prediction of macroscopic properties.

The proposed novel concept of spatially mapping latent space features is highly applicable to metal microstructure characterization, as it maintains the spatial representation of heterogeneities found in metallic materials, which is a key factor in determining macroscopic properties.

We discuss identification of microstructure heterogeneity. It is observed from Figs. 4–7 that the proposed approach, which encodes and maps data within the latent space, enables the identification of microstructural heterogeneities at various scales beyond conventional EBSD analysis. When applied to a wrought alloy with fully recrystallized grains, the grain structure is revealed, and the contrast observed is mostly related to the crystallographic orientation, specifically the position of the bands within the Kikuchi pattern. Results compare to conventional methods, such as the Hough transform[24] or dictionary-based approaches[25,87] that rely on the band positions and comparing them with theoretical patterns. Nevertheless, as shown in Fig. 4B, the proposed encoding and mapping approach reveals a small twin (black arrow in Fig. 4B) that is not visible in the IPF maps presented in Fig. 2A. Despite using the same experimental measurements, the encoding and mapping approach highlights smaller-scale features, thereby providing enhanced sensitivity in identifying microstructural heterogeneities. This enhancement can be explained by the nature of EBSD measurements, where the measurement is not discrete but relies on averaging the diffraction information over the interaction volume[88]. When the measurement is performed near grain boundaries, the interaction volume often encompasses two distinct crystallographic orientations, leading to the overlapping of Kikuchi patterns. Conventional EBSD analysis assumes the presence of only a single pattern, ignoring such overlaps. However, in the case of encoding Kikuchi patterns, the latent space features derived from overlapping patterns differ significantly from those obtained for a single diffraction pattern. As a result, the proposed approach achieves better spatial

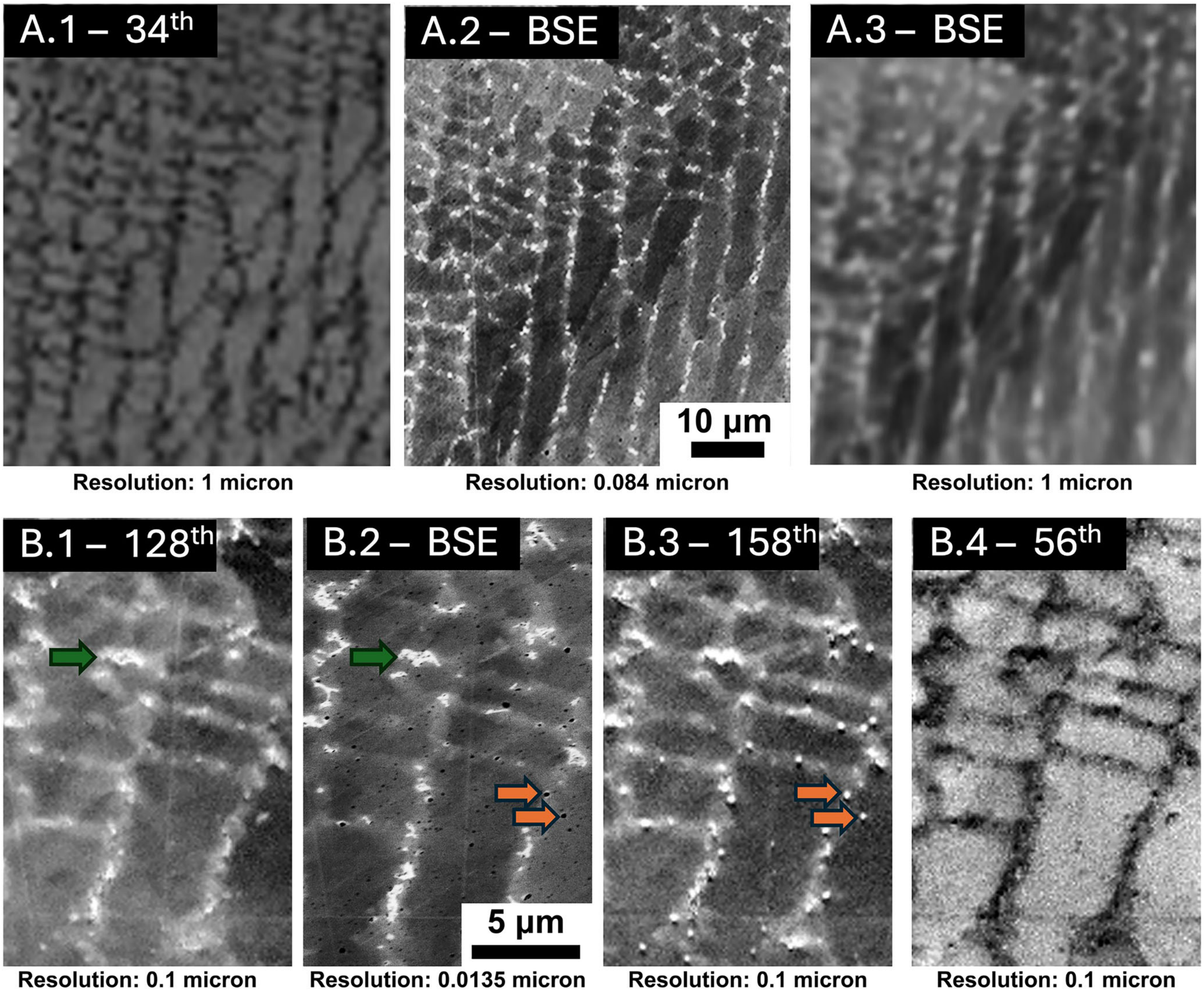

**Fig. 12 | Revealing chemistry and dislocation density in AM materials. A.1** Representation of the 34th latent space feature map for a reduced region of interest in the AM as-built material. Spatial resolution during acquisition was 1 μm (i.e., 1 μm EBSD step size). **A.2, A.3** Associated BSE images with spatial resolution of 84 nm and 1 μm, respectively. **B.1, B.3, B.4** High-resolution representation of the 128, 158 and 56th latent space feature maps for a reduced region of interest. Spatial resolution during acquisition was 0.1 μm (i.e., 0.1 μm EBSD step size). **(B.2)** The associated BSE image.

sensitivity in detecting microstructural heterogeneities with sub-interaction volume resolution.

The efficiency of the proposed method in identifying microstructural heterogeneities is particularly evident when multiple types of heterogeneities, occurring at different scales, are present within a material. This is especially true for AM materials, where chemical segregation, small particles, cellular structures, lattice rotation and expansion, and crystallographic orientation are commonly observed[80,84]. While all these features are inherently present in the diffraction patterns, they are challenging to extract physically and even more difficult to distinguish from one another. For instance, cellular structures, often associated with variations in dislocation density, are not detectable using conventional EBSD analysis. The only existing approach is the sharpness analysis[11], which quantifies the diffuseness of Kikuchi bands and theoretically relates to statistically stored dislocation density[68]. However, this method typically requires high-resolution Kikuchi patterns to achieve sufficient sensitivity for detecting cellular structures in AM materials[11]. Sharpness analysis applied to the collected Kikuchi patterns provided limited identification of cellular structures, as shown in Fig. 5B.5, C.5. When performed with high spatial resolution (Fig. 7B, D, the cellular structures were more apparent but still lacked the sensitivity observed with the proposed encoding and mapping method, as shown in Fig. 7A.1, C.1. Similar findings apply to small lattice rotations and expansions, which typically require advanced methods such as high-resolution EBSD (HR-EBSD) for detection[27]. For example, Fig. 7C.2 highlights subdomains (contrast variations) within cells, corresponding to small lattice expansions and rotations induced by the AM process[80]. Interestingly, the proposed method also demonstrates effectiveness in detecting small particles located at the cell walls, as illustrated in Fig. 7C.3. Conventional EBSD methods lack the capability to differentiate phases within a diffraction pattern without specific analysis.

Other characterization techniques, such as electron channeling contrast imaging (ECCI)[82,89,90], transmission electron microscopy (TEM), or conventional high-resolution backscattered electron (BSE) imaging, can provide insight into the fine-scale microstructure of AM materials. As an example, we compare conventional BSE images with the latent space feature maps extracted from our encoder in Fig. 12, at two different spatial resolutions (1 and 0.1 μm). While BSE imaging allows observation of chemical heterogeneities, it is limited in capturing dislocation structures unless channeling contrast is used, a time-consuming process requiring multiple tilt angles, particularly for AM and polycrystalline materials. In the conventional BSE image acquired at 1 µm resolution (Fig. 12A.3), cell walls are not clearly resolved, unlike in the 34th feature map (Fig. 12A.1), highlighting the enhanced sensitivity and spatial resolution of the proposed encoding method. At higher resolution (0.1 µm, Fig. 12B), the 128th feature map shows strong sensitivity not only to chemical segregation along cell walls but also to the presence of bright particles (indicated by green arrows in Fig. 12B.1, B.2). Interestingly, the 158th feature map captures both the dark particles and the chemical segregation (orange arrows in Fig. 12B.2, B.3). The proposed approach not only captures different phases and chemical segregation patterns but also distinguishes them within the latent space representation.

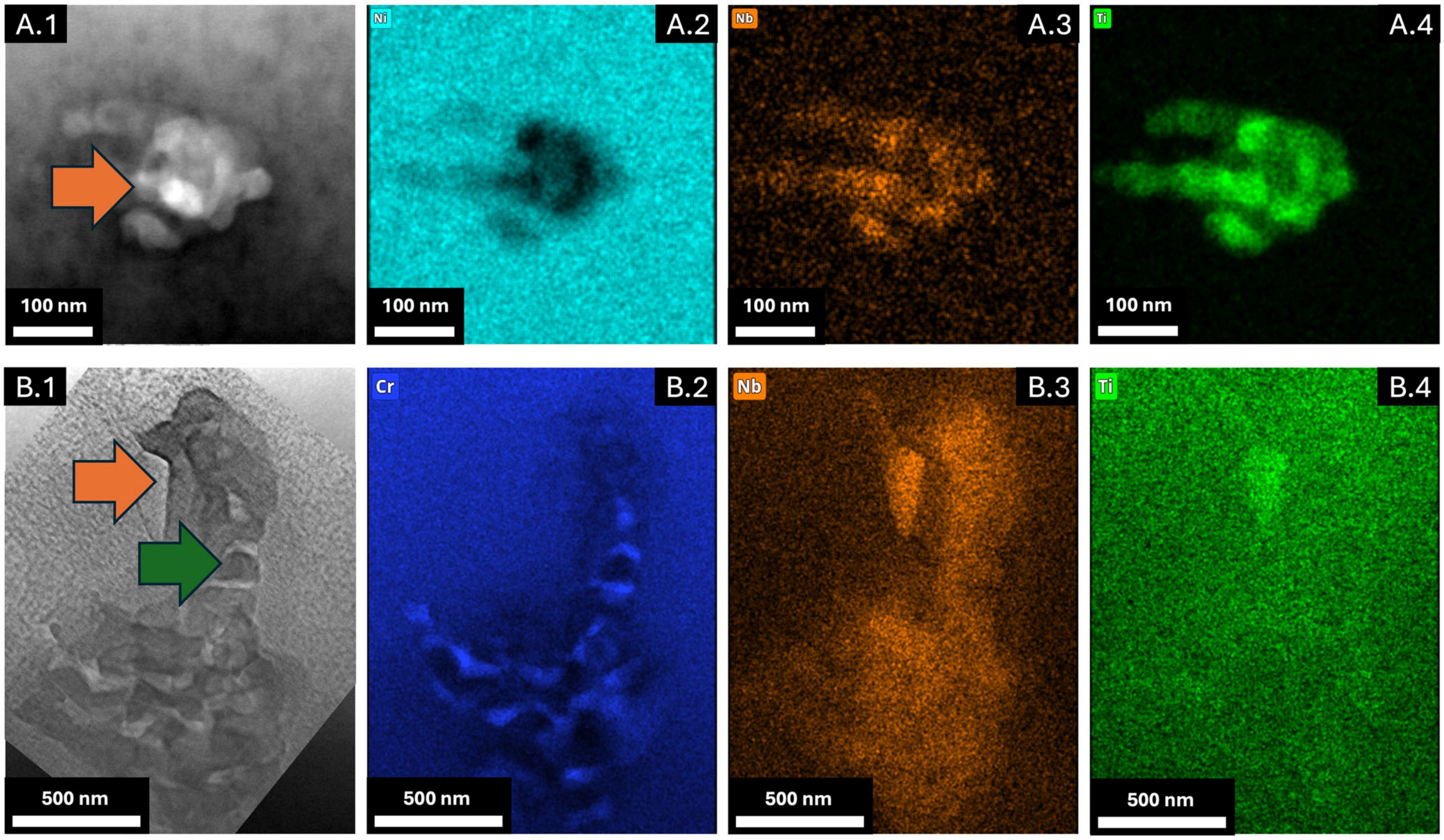

**Fig. 13 | Particles in AM Inconel 718. (A.1, B.1)** STEM images of the two distinct types of particles (see green and orange arrows in Fig. 12) identified in the investigated AM as-built materials. **(A.2-A.4, B.2-B.4)** Corresponding EDX elemental maps.

To further validate that variations in composition and structure contribute differently to the latent feature maps, we identified the nature of the bright particles observed in Fig. 12B.1–B.3. A thin foil was extracted from the same region where high-resolution EBSD was performed. EDX elemental mapping, presented in Fig. 13, confirms distinct chemical compositions for the particles indicated by green and orange arrows in Fig. 12B.1–B.3. The smaller particles (orange arrows, shown in Fig. 13A) are likely Nb- and Ti-rich phases, consistent with Nb- and/or Ti-rich MC carbides[69], while the larger particles are likely Nb-rich Laves phases[69].

While sensitivity is significantly increased through encoding Kikuchi patterns and mapping the latent space features, physical interpretation of individual latent space features is lost, owing to the way VAEs encode information into latent spaces. A single latent space feature may represent a non-linear combination of multiple physical microstructural characteristics (i.e., crystallographic orientation, dislocation density, chemical composition, lattice expansion, phase) as well as non-physical factors (i.e., detector geometry or experimental noise). However, it is observed that increasing dimensionality within the latent space provides an effective solution in the disentanglement of the physical features as shown in Fig. 9. Encoding data into a much higher dimensional latent space may introduce more and more features related to non-physical information, such as noise or artifacts from the experimental setup. Determining the optimal latent space dimension for visualizing and quantifying microstructural heterogeneities remains an open question. Another promising approach to identify the physical content represented in each latent space map is the application of interpretability tools, such as activation mapping[91], occlusion sensitivity analysis[92], Grad-CAM[93,94], LIME[95,96], and permutation importance. These methods can help localize the regions within Kikuchi patterns that predominantly influence each latent feature, enabling a physical interpretation of their significance. For example, if a latent space feature is primarily associated with the edges of Kikuchi bands, it may indicate that dislocation density is a key contributor to that feature.

We discuss encoding and mapping for data-based prediction. Another advantage of the proposed approach, beyond identifying microstructural heterogeneities, is its potential to reduce complex microstructures into a latent space representation suitable for use in data-driven prediction models. Considering the 256- or 128-dimensional latent space, the proposed approach reduces the data volume by a factor of 900 and 1800, respectively. In addition, this method simplifies microstructure representation by directly encoding diffraction data without any prior knowledge or annotations needed. Traditional approaches to simplify microstructure representations often rely on graph neural networks[39,41–43,97,98], which require an initial physical preprocessing step, such as grain segmentation. These methods typically utilize simple physical descriptors of microstructure, such as the average grain orientation, grain size, and geometry. However, for complex microstructures, such as those produced by AM, these traditional methods are unable to account for features like cellular structures, lattice expansion, and chemical fluctuations. In contrast, the present method theoretically preserves all the information contained in the diffraction patterns by encoding the raw data directly. The sensitivity is significantly enhanced compared to conventional or advanced physical EBSD processing, as demonstrated in Figs. 5–7. These figures reveal fluctuations related to lattice expansion, dislocation density, and crystallographic orientation and structure that are not detectable using conventional or advanced EBSD physics-based processing methods. This demonstrates that by encoding metal diffraction data and mapping the associated latent space features, we can extract all microstructure features without relying on physical assumptions or prior knowledge of the existing features.

**Table 2 | Detailed architectures of CNNs used for Kikuchi patterns encoding/decoding**

| | Output size | Kernel size | 16 | 32 | 64 | 128 | 256 |
|---|---|---|---|---|---|---|---|
| Encoder | 56 × 56 | 9 × 9, stride 2 | 4 | 8 | 16 | 32 | 64 |
| | 56 × 56 | 9 × 9 | 4 (*) | 8 (*) | 16 (*) | 32 (*) | 64 (*) |
| | 24 × 24 | 9 × 9, stride 2 | 8 | 16 | 32 | 64 | 128 |
| | 24 × 24 | 9 × 9 | 8 (*) | 16 (*) | 32 (*) | 64 (*) | 128 (*) |
| | 8 × 8 | 9 × 9, stride 2 | 16 | 32 | 64 | 128 | 256 |
| | 8 × 8 | 9 × 9 | 16 (*) | 32 (*) | 64 (*) | 128 (*) | 256 (*) |
| | 1 × 1 | 8 × 8 | 32 | 64 | 128 | 256 | 512 |
| | 1 × 1 | / | 32-D fc | 64-D fc | 128-D fc | 256-D fc | 512-D fc |
| | / | / | sampling layer | | | | |
| Parameters | / | / | 66k | 297k | 1185k | 4738k | 18,943k |
| Decoder | 4 × 4 | / | projection and reshape | | | | |
| | 11 × 11 | 9 × 9, stride 2 | 16 | 32 | 64 | 128 | 256 |
| | 11 × 11 | 9 × 9 | 16 (*) | 32 (*) | 64 (*) | 128 (*) | 256 (*) |
| | 27 × 27 | 9 × 9, stride 2 | 8 | 16 | 32 | 64 | 128 |
| | 27 × 27 | 9 × 9 | 8 (*) | 16 (*) | 32 (*) | 64 (*) | 128 (*) |
| | 61 × 61 | 9 × 9, stride 2 | 4 | 8 | 16 | 32 | 64 |
| | 61 × 61 | 9 × 9 | 4 (*) | 8 (*) | 16 (*) | 32 (*) | 64 (*) |
| | 120 × 120 | 9 × 9, stride 2 | 2 | 4 | 8 | 16 | 32 |
| | 120 × 120 | 9 × 9 | 2 (*) | 4 (*) | 8 (*) | 16 (*) | 32 (*) |
| | 120 × 120 | 1 × 1 | 1 | 1 | 1 | 1 | 1 |
| Parameters | / | / | 74k | 265k | 1057k | 4225k | 16,896k |

*Indicates additional details about the architecture.

Recent studies have demonstrated that spatial encoding of plasticity using similar VAE and latent space feature mapping approaches can accurately predict macroscopic mechanical properties[99], highlighting the importance of accounting for material heterogeneity. The inclusion of the encoded microstructure (as proposed here) in such predictive frameworks may therefore improve the accuracy of macroscopic property prediction by capturing microstructural heterogeneity. More importantly, integrating microstructural features and their heterogeneity opens the possibility of identifying microstructure effects and designing microstructures directly within the latent space, thereby establishing a new research direction in alloy design.

When using this method as input for any data-driven model, the disentanglement of physical features is not so critical. Instead, the focus shifts to designing the latent space to provide a better and more continuous representation of all possible Kikuchi patterns to further increase sensitivity. If the latent space is overly clustered, the model may struggle to differentiate between closely related diffraction states. Therefore, optimizing the latent space for smooth, comprehensive coverage of diffraction patterns is essential for achieving accurate and robust predictions in data-based models.

Another key aspect of achieving smooth and comprehensive coverage in the latent space (a continuous latent space) is ensuring that all feature values within the latent space correspond to realistic Kikuchi patterns. This is essential for effective design and functionality within the latent space. With a continuous latent space, microstructures can be generated directly, allowing material optimization to occur within the latent space itself.

We discuss structure of latent space. The framework proposed in this study demonstrated its ability to produce a low-dimensional representation of Kikuchi patterns and to capture key features contained in the diffraction patterns. However, the structure of the latent space depends strongly on several hyperparameters considered in this study, such as the number of dimensions of the latent space, the structure of both the *Encoder* and the *Decoder*, but also the loss function used during training. All these parameters significantly affect the reconstruction quality of Kikuchi patterns as demonstrated in Table 1. Increasing the number of dimensions improves the quality of the reconstruction, but also favors the identification of microstructural features such as the cellular structure shown in Fig. 9. 256 dimensions, the largest number of latent space dimensions considered in this study produces a reconstruction similar to the original patterns. Ultimately, using a much larger number of latent space dimensions will improve reconstruction, but may also find structure in the noise of the Kikuchi patterns and produce feature maps that are no longer representative of the microstructure under consideration. Different loss functions were employed in addition to conventional VAEs. Despite the use of an additional contrastive loss, performances are found to be similar to the conventional 128-dimension VAE (Table 1). On the other hand, increasing the weight of the Kullback-Leibler divergence by using $\beta$-VAE greatly affects the quality of the reconstructions. While some reconstructed patterns are of excellent quality, others do not show any bands (Fig. 10B.2) or are erroneously reconstructed (Fig. 10C.2). In Higgins et al.[70], the $\beta$-VAE learned a more efficient low-dimensional representation on datasets such as faces[100] and chairs[101]. More generally, increasing $\beta$ should help the model to align the latent dimensions with the various factors found in the image[102], but has also been reported to globally affect the quality of the reconstruction[103]. For Kikuchi patterns, most of the underlying factors of variation are not independent and depend on the crystal reciprocal space. Some other factors of variation, such as the pattern distortion[87] or dislocation density[11], could be decoupled but represent smaller variations in a pattern compared to the crystallographic orientation. Beyond affecting the quality of reconstruction, increasing $\beta$ leads to erroneous reconstructions, as shown in Fig. 10, analogous to the 'mode collapse' observed for Generative Adversarial Networks (GAN)[104]. The proposed approaches reveal a trade-off between accurately reconstructing Kikuchi patterns, crucial for identifying microstructure heterogeneity, and achieving a continuous latent space capable of predicting realistic Kikuchi patterns to guide the design of new microstructures. One solution involves leveraging a conditional VAE-GAN architecture trained on synthetic data generated through forward modeling[105]. This combination of VAE and GAN networks allows for the modification of the latent space structure, as the discriminator network evaluates both training diffraction patterns and patterns randomly sampled from the latent space during training. Meanwhile, it is worth reminding that the low-dimensional representations no longer have a direct physical meaning due to the way VAEs encode information into latent spaces. This directly influences the visualization and identification of heterogeneities. For example, cellular structures shown in Fig. 4C and Fig. 4 are observed on different latent space maps with respect to the crystallographic orientation of the grains of interest. To improve dimension disentanglement and to facilitate identification of heterogeneities, a rotation-invariant VAE could be employed[106]. This way, training could be performed to target specific physical features.

We discuss perspectives. The proposed encoding methods have the potential to efficiently reveal heterogeneities and could be extended to other material characterization techniques, such as EDS. Multi-modal latent feature maps (combining encoded EBSD and EDS for instance) can provide a solution to increase sensitivity to different types of heterogeneities (chemical, dislocation, orientation, phase) and/or at different scales[107]. The content of the latent space could be enhanced toward certain heterogeneities by means of semi-supervised or supervised learning (conditional statements and/or classifiers)[105]. Moreover, the mapped low-dimensional representation can be further used for heterogeneity identification and microstructure segmentation without physics-based analysis[108]. It creates opportunities for autonomous identification of microstructural features and their heterogeneity, paving the way for uncovering the microstructure genome.

Furthermore, the present approach enables the encoding of metallic microstructures in a form suitable for machine learning-based predictive and design models. This opens the field of "Material Spatial Intelligence", which leverages data-driven frameworks that integrate spatially resolved characterization data. Unlike traditional methods that rely on average properties, material spatial intelligence captures microstructural heterogeneity across multiple length scales to more accurately understand and predict material behavior. By embedding this rich microstructural information into predictive models, the approach may enable rapid and accurate property estimation, inverse design, and deeper insights into structure-property relationships, ultimately accelerating alloy development and performance optimization.

## METHODS

### Materials

Three different nickel-based superalloys were used in this study, a wrought recrystallized Inconel 718 (30 min at 1050 °C followed by 8 h at 720 °C) with chemical composition of (wt.%) Ni - 0.56% Al - 17.31% Fe - 0.14% Co - 17.97% Cr - 5.4% Nb + Ta - 1.00% Ti - 0.023% C - 0.0062% N; a 3D-printed Inconel 718 by directed energy deposition (DED) (as-built) ; a partially recrystallized Waspalloy (heat-treated) characterized by a necklace microstructure (for a training purpose but not detailed in the present study). The 3D-printed material was produced using a Formalloy L2 DED unit utilizing a 650 W Nuburu 450 nm blue laser capable of achieving a 400 μm laser spot size. Argon was used as the shielding and carrier gas, and the specimen remained in its as-built condition. The chemical composition is in wt.%, Ni - 0.45% Al - 18.77% Fe - 0.07% Co - 18.88% Cr - 5.08% Nb - 0.96% Ti - 0.036% C - 0.02% Cu - 0.04% Mn - 0.08% Si - 3.04% Mo. All samples were machined by EDM as flat dogbone samples of gauge section $1 \times 3$ mm$^2$. All samples were mechanically polished using abrasive papers followed by diamond suspension down to 3 μm and were finished using a 50 nm colloidal silica suspension.

### Electron backscatter diffraction

EBSD measurements were performed on a ThermoFischer Scios 2 Dual Beam SEM/FIB with an EDAX OIM-Hikari detector with a 1 and 0.1 μm step size, at an accelerating voltage 20 kV, current of 6.4 nA, an exposure time of 8.5 ms, 12 mm of working distance and 70° tilt. In total, 6 maps of $1000 \times 900$ μm were collected. Kikuchi patterns with no binning were saved to 16-bit images under the UP2 format. They were finally extracted using the EDAX OIM 8 software with no background correction.

### Back scatter diffraction electron microscopy and transmission electron microscopy

A ThermoFisher Scientific Helios Nanolab 600i SEM-FIB Dual-Beam system equipped with a high-angle backscattered electron (BSE) detector was used to capture BSE images of the investigated regions. Imaging was performed at an accelerating voltage of 5 kV and a beam current of 0.34 nA, with a working distance of 6 mm. The spatial resolution of the acquired images was either 13 nm, 84 nm, or 1 μm. The same equipment was then used to extract the thin foil with a final cleaning at 5 kV and a current of 16 pA. A ThermoFisher Scientific Talos F200X G2 Scanning Transmission Electron Microscope (S/TEM) was used to capture conventional STEM images. EDS was then performed at 200 kV.

### VAE architecture

The complete VAE architecture is detailed in Fig. 3C, composed of two CNNs, *Encoder* and *Decoder* portions, to project Kikuchi patterns to a low-dimensional representation space and to restore the original pattern from this low-dimensional representation. The *Encoder* and *Decoder* consist of large convolution kernels to limit the depth of the network and are detailed in Table 2 for the different latent space dimensions. Leaky ReLU with a scale factor of 0.2 were used as activation functions after each convolution layer[109] (and transposed convolution), except for the output layer of the decoder network, for which a hyperbolic tangent was used. The *Encoder* terminates by a fully connected layer followed by a sampling layer[62]. To accelerate the training, "Shortcut connections" – popularized by ResNets[67] – were used on layers marked as * in Table 2. We conducted a parametric study (not presented here) by systematically increasing the network depth and the number of parameters within the architecture, and observed no significant improvement in performance, highlighting the robustness and efficiency of the proposed approach even with a low-depth network. All these architectures were implemented on MATLAB using the Deep Learning Toolbox. An example of VAE architecture including both reconstruction loss and KL divergence is given in ref. 110.

### Data preprocessing and VAE training

Before VAE training, Kikuchi pattern backgrounds were calculated over each EBSD map to be individually subtracted from each pattern. To reduce the number of CNN parameters, a binning of 4 was applied. The patterns were also rescaled to the $[-1, 1]$ range to match the scale of hyperbolic tangent output function. To encode the Kikuchi patterns, various dimensions of latent spaces were considered ranging from 16 to 256 values. All VAEs have been trained with a total of 96k patterns, 16k patterns randomly selected within each EBSD map (before and after mechanical testing). The dataset has been randomly divided into two portions, 80% for training and 20% for validation. All the CNN architectures have been trained with the Adam updater[111] and using the same parameters, a batch size of 256 for 250 epochs, learning rate of 0.001, gradient decay of 0.9 and squared gradient decay of 0.999. The trainings took about 6 hours regardless of the latent space dimensions and about 40 hours with the augmentations for the adapted methodology for contrastive learning[63]. All trainings have been performed using the same CPU and GPU configuration, Intel i7-14700F and NVIDIA Geforce RTX 4090.

## Data availability

Binned and preprocessed data are available through the Dryad Digital Repository (https://doi.org/10.5061/dryad.zcrjdfnr9). The full-resolution raw data will be provided upon request to the corresponding author. The architectures used in this study were implemented on MATLAB and the associated code will be made available upon request to the corresponding author.

## Code availability

The architectures used in this study were implemented on MATLAB and the associated code will be made available upon request to the corresponding author.

## References

1. Stinville, J., Vanderesse, N., Bridier, F., Bocher, P. & Pollock, T. High resolution mapping of strain localization near twin boundaries in a nickel-based superalloy. *Acta Materialia* **98**, 29–42 (2015).
2. Charpagne, M. et al. Slip localization in inconel 718: a three-dimensional and statistical perspective. *Acta Materialia* **215**, 117037 (2021).
3. Hémery, S. et al. Strain localization and fatigue crack formation at (0001) twist boundaries in titanium alloys. *Acta Materialia* **219**, 117227 (2021).
4. Echlin, M. P., Stinville, J. C., Miller, V. M., Lenthe, W. C. & Pollock, T. M. Incipient slip and long range plastic strain localization in microtextured ti-6al-4v titanium. *Acta Materialia* **114**, 164–175 (2016).
5. Edwards, T. E. J., Maeder, X., Ast, J., Berger, L. & Michler, J. Mapping pure plastic strains against locally applied stress: revealing toughening plasticity. *Sci. Adv.* **8**, eabo5735 (2022).
6. Harte, A. et al. The effect of solid solution and gamma prime on the deformation modes in ni-based superalloys. *Acta Materialia* **194**, 257–275 (2020).
7. Zhang, Z. et al. Strain localisation and failure at twin-boundary complexions in nickel-based superalloys. *Nat. Commun.* **11**, 4890 (2020).
8. Hu, H., Briffod, F., Yin, W., Shiraiwa, T. & Enoki, M. Quantitative investigation of slip band activities in a bimodal titanium alloy under pure fatigue and dwell-fatigue loadings. *Int. J. Fatigue* **182**, 108203 (2024).
9. Bean, C. et al. Heterogeneous slip localization in an additively manufactured 316l stainless steel. *Int. J. Plasticity* **159**, 103436 (2022).
10. Abdi, F. et al. Grain boundary engineering of new additive manufactured polycrystalline alloys. *Forces Mech.* **4**, 100033 (2021).
11. Wang, F. et al. Dislocation cells in additively manufactured metallic alloys characterized by electron backscatter diffraction pattern sharpness. *Mater. Charact.* **197**, 112673 (2023).
12. Zhou, K. et al. A precipitation-strengthened high-entropy alloy for additive manufacturing. *Addit. Manuf.* **35**, 101410 (2020).
13. Hu, D. et al. Understanding the strain localization in additively manufactured materials: micro-scale tensile tests and crystal plasticity modeling. *Int. J. Plasticity* **177**, 103981 (2024).
14. Wang, A. Y.-T., Kauwe, S. K., Murdock, R. J. & Sparks, T. D. Compositionally restricted attention-based network for materials property predictions. *npj Comput. Mater.* **7**, 77 (2021).
15. Mansouri Tehrani, A., Ghadbeigi, L., Brgoch, J. & Sparks, T. D. Balancing mechanical properties and sustainability in the search for superhard materials. *Integr. Mater. Manuf. Innov.* **6**, 1–8 (2017).
16. Gianola, D. S. et al. Advances and opportunities in high-throughput small-scale mechanical testing. *Curr. Opin. Solid State Mater. Sci.* **27**, 101090 (2023).
17. Marano, A., Ribart, C. & Proudhon, H. Towards a data platform for multimodal 4D mechanics of material microstructures. *Mater. Des.* **246**, 113306 (2024).
18. Ziatdinov, M. et al. Deep learning of atomically resolved scanning transmission electron microscopy images: chemical identification and tracking local transformations. *ACS Nano* **11**, 12742–12752 (2017).
19. Yang, B., Vassilev-Galindo, V. & Llorca, J. Application of machine learning to assess the influence of microstructure on twin nucleation in Mg alloys. *npj Comput. Mater.* **10**, 26 (2024).
20. Courtright, Z. S. et al. High-throughput experiments and machine learning strategies for efficient exploration of additively manufactured inconel 625. *Acta Materialia* **288**, 120875 (2025).
21. Adapa, V. S. K., Kalidindi, S. R. & Saldana, C. J. Rapid development of metal additive manufacturing using artificial intelligence/machine learning and high-throughput material testing. *Annu. Rev. Mater. Res.* **55**, 175–201 (2025).
22. Murgas, B., Stickel, J., Brewer, L. & Ghosh, S. Modeling complex polycrystalline alloys using a generative adversarial network enabled computational platform. *Nat. Commun.* **15**, 9441 (2024).
23. Ghosh, S., Dimiduk, D. & Furrer, D. Statistically equivalent representative volume elements (serve) for material behaviour analysis and multiscale modelling. *Int. Mater. Rev.* **68**, 1158–1191 (2023).
24. Duda, R. O. & Hart, P. E. Use of the Hough transformation to detect lines and curves in pictures. *Commun. ACM* **15**, 11–15 (1972).
25. Wright, S. I. et al. Introduction and comparison of new EBSD post-processing methodologies. *Ultramicroscopy* **159**, 81–94 (2015).
26. Lenthe, W., Singh, S. & De Graef, M. A spherical harmonic transform approach to the indexing of electron back-scattered diffraction patterns. *Ultramicroscopy* **207**, 112841 (2019).
27. Wilkinson, A. J., Meaden, G. & Dingley, D. J. High-resolution elastic strain measurement from electron backscatter diffraction patterns: new levels of sensitivity. *Ultramicroscopy* **106**, 307–313 (2006).
28. Winkelmann, A. et al. Assignment of chiral elemental crystal structures using kikuchi diffraction. *Mater. Charact.* **196**, 112633 (2023).
29. Gallaugher, M., Brodusch, N., Gauvin, R. & Chromik, R. R. Magnetic domain structure and crystallographic orientation of electrical steels revealed by a forescatter detector and electron backscatter diffraction. *Ultramicroscopy* **142**, 40–49 (2014).
30. Tanaka, T., Kawakami, K., Mogi, H. & Arai, S. An EBSD camera as a tool to characterise in-plane magnetisation vectors on FE-SI (001) surface. *Ultramicroscopy* **250**, 113754 (2023).
31. Naresh-Kumar, G. et al. Quantitative imaging of anti-phase domains by polarity sensitive orientation mapping using electron backscatter diffraction. *Sci. Rep.* **7**, 10916 (2017).
32. Konijnenberg, P., Zaefferer, S. & Raabe, D. Assessment of geometrically necessary dislocation levels derived by 3d ebsd. *Acta Materialia* **99**, 402–414 (2015).
33. Schwartz, A. J., Kumar, M., Adams, B. L. & Field, D. P. *Electron backscatter diffraction in materials science*, vol. 2 (Springer, 2009).
34. Wilkinson, A. J., Collins, D. M., Zayachuk, Y., Korla, R. & Vilalta-Clemente, A. Applications of multivariate statistical methods and simulation libraries to analysis of electron backscatter diffraction and transmission kikuchi diffraction datasets. *Ultramicroscopy* **196**, 88–98 (2019).
35. Bonnet, N. Multivariate statistical methods for the analysis of microscope image series: applications in materials science. *J. Microsc.* **190**, 2–18 (1998).
36. Brewer, L. N., Kotula, P. G. & Michael, J. R. Multivariate statistical approach to electron backscattered diffraction. *Ultramicroscopy* **108**, 567–578 (2008).
37. Wright, S. I., Nowell, M. M., de Kloe, R., Camus, P. & Rampton, T. Electron imaging with an EBSD detector. *Ultramicroscopy* **148**, 132–145 (2015).
38. Brodusch, N., Demers, H. & Gauvin, R. Imaging with a commercial electron backscatter diffraction (ebsd) camera in a scanning electron microscope: a review. *J. Imaging* **4**, 88 (2018).
39. Dai, M., Demirel, M. F., Liang, Y. & Hu, J.-M. Graph neural networks for an accurate and interpretable prediction of the properties of polycrystalline materials. *npj Comput. Mater.* **7**, 103 (2021).
40. Qin, Y., DeWitt, S., Radhakrishnan, B. & Biros, G. Graingnn: a dynamic graph neural network for predicting 3D grain microstructure. *J. Comput. Phys.* **510**, 113061 (2024).
41. Thomas, A. et al. Materials fatigue prediction using graph neural networks on microstructure representations. *Sci. Rep.* **13**, 12562 (2023).
42. Hestroffer, J. M., Charpagne, M.-A., Latypov, M. I. & Beyerlein, I. J. Graph neural networks for efficient learning of mechanical properties of polycrystals. *Comput. Mater. Sci.* **217**, 111894 (2023).

43. Pagan, D. C., Pash, C. R., Benson, A. R. & Kasemer, M. P. Graph neural network modeling of grain-scale anisotropic elastic behavior using simulated and measured microscale data. *npj Comput. Mater.* **8**, 1–12 (2022).
44. Pandey, A. & Pokharel, R. Machine learning based surrogate modeling approach for mapping crystal deformation in three dimensions. *Scr. Materialia* **193**, 1–5 (2021).
45. Mangal, A. & Holm, E. A. Applied machine learning to predict stress hotspots i: face centered cubic materials. *Int. J. Plasticity* **111**, 122–134 (2018).
46. Adams, B. L., Gao, X. C. & Kalidindi, S. R. Finite approximations to the second-order properties closure in single phase polycrystals. *Acta Materialia* **53**, 3563–3577 (2005).
47. Cecen, A., Fast, T. & Kalidindi, S. R. Versatile algorithms for the computation of 2-point spatial correlations in quantifying material structure. *Integrating Mater. Manuf. Innov.* **5**, 1–15 (2016).
48. Paulson, N. H., Priddy, M. W., McDowell, D. L. & Kalidindi, S. R. Reduced-order structure-property linkages for polycrystalline microstructures based on 2-point statistics. *Acta Materialia* **129**, 428–438 (2017).
49. Knezevic, M. & Kalidindi, S. R. Crystal plasticity modeling of microstructure evolution and mechanical fields during processing of metals using spectral databases. *JOM* **69**, 830–838 (2017).
50. Jung, H. G. & Kim, H. S. Exploration of optimal microstructure and mechanical properties in continuous microstructure space using a variational autoencoder. *Mater. Des.* **202**, 109544 (2021).
51. Hu, C., Martin, S. & Dingreville, R. Accelerating phase-field predictions via recurrent neural networks learning the microstructure evolution in latent space. *Comp. Methods Appl. Mech. Eng.* **397**, 115128 (2022).
52. Jones, R. E. et al. Multiscale simulation of spatially correlated microstructure via a latent space representation. *Int. J. Solids Struct.* **301**, 112966 (2024).
53. McAuliffe, T., Dye, D. & Britton, T. Spherical-angular dark field imaging and sensitive microstructural phase clustering with unsupervised machine learning. *Ultramicroscopy* **219**, 113132 (2020).
54. Valleti, M., Ziatdinov, M., Liu, Y. & Kalinin, S. V. Physics and chemistry from parsimonious representations: image analysis via invariant variational autoencoders. *npj Comput. Mater.* **10**, 183 (2024).
55. Generale, A. P., Robertson, A. E., Kelly, C. & Kalidindi, S. R. Inverse stochastic microstructure design. *Acta Materialia* **271**, 119877 (2024).
56. Li, S.-H., Kumar, P., Chandra, S. & Ramamurty, U. Directed energy deposition of metals: processing, microstructures, and mechanical properties. *Int. Mater. Rev.* **68**, 605–647 (2023).
57. Cabeza, S. et al. Strain monitoring during laser metal deposition of inconel 718 by neutron diffraction. In *Superalloys 2020: Proceedings of the 14th International Symposium on Superalloys*, 1033–1045 (Springer, 2020).
58. Mueller, M. et al. Influence of process parameter variation on the microstructure of thin walls made of Inconel 718 deposited via laser-based directed energy deposition with blown powder. *J. Mater. Sci.* **58**, 11310–11326 (2023).
59. Nie, Y., Chang, Y. & Charpagne, M. Origins of twin boundaries in additive manufactured stainless steels. *Acta Materialia* **275**, 120035 (2024).
60. Nie, Y., Chang, Y. & Charpagne, M. Functionally graded stainless steels with tailored grain boundary serration. *Scr. Materialia* **237**, 115714 (2023).
61. Calvat, M., Anjaria, D., Wang, H., Vecchio, K. & Stinville, J. Kikuchi pattern dataset from wrought and as-built additively manufactured superalloys. (Dryad Digital Repository, 2025).
62. Kingma, D. P. Auto-encoding variational bayes. *arXiv preprint arXiv:1312.6114* (2013).
63. Chen, T., Kornblith, S., Norouzi, M. & Hinton, G. A simple framework for contrastive learning of visual representations (2020). arXiv: 2002.05709.
64. Ding, Z., Pascal, E. & De Graef, M. Indexing of electron back-scatter diffraction patterns using a convolutional neural network. *Acta Materialia* **199**, 370–382 (2020).
65. Doersch, C. Tutorial on variational autoencoders. *arXiv preprint arXiv:1606.05908* (2016).
66. Oord, A. v. d., Li, Y. & Vinyals, O. Representation learning with contrastive predictive coding. *arXiv preprint arXiv:1807.03748* (2018).
67. He, K., Zhang, X., Ren, S. & Sun, J. Deep residual learning for image recognition. In *Proc. IEEE Conference on Computer Vision and Pattern Recognition*, 770–778 (IEEE, 2016).
68. Zhu, C. & De Graef, M. Ebsd pattern simulations for an interaction volume containing lattice defects. *Ultramicroscopy* **218**, 113088 (2020).
69. Bambach, M., Sizova, I., Kies, F. & Haase, C. Directed energy deposition of inconel 718 powder, cold and hot wire using a six-beam direct diode laser set-up. *Addit. Manuf.* **47**, 102269 (2021).
70. Higgins, I. et al. Beta-vae: learning basic visual concepts with a constrained variational framework. In *Proc. ICLR (Poster)***3** (ICLR, 2017).
71. Kikuchipy (2024). Accessed: 2025-06-20.
72. Bachmann, F., Hielscher, R. & Schaeben, H. Texture analysis with mtex–free and open source software toolbox. *Solid State Phenom.* **160**, 63–68 (2010).
73. Foden, A., Previero, A. & Britton, T. B. Advances in electron backscatter diffraction. *arXiv preprint arXiv:1908.04860* (2019).
74. Kaufmann, K. et al. Crystal symmetry determination in electron diffraction using machine learning. *Science* **367**, 564–568 (2020).
75. Kaufmann, K. et al. Phase mapping in ebsd using convolutional neural networks. *Microsc. Microanal.* **26**, 458–468 (2020).
76. Xiong, G., Wang, C., Yan, Y., Zhang, L. & Su, Y. Indexing high-noise electron backscatter diffraction patterns using convolutional neural network and transfer learning. *Comput. Mater. Sci.* **233**, 112718 (2024).
77. Chauniyal, A., Thome, P. & Stricker, M. Employing constrained nonnegative matrix factorization for microstructure segmentation. *Microsc. Microanal.* **30**, 712–723 (2024).
78. Lu, Q. et al. Crystal orientation and deformation state analysis from kikuchi patterns via pattern reconstruction aided deep siamese network. *Mater. Des.* **230**, 111998 (2023).
79. Andrews, C. E., Strantza, M., Calta, N. P., Matthews, M. J. & Taheri, M. L. A denoising autoencoder for improved kikuchi pattern quality and indexing in electron backscatter diffraction. *Ultramicroscopy* **253**, 113810 (2023).
80. Wang, Y. M. et al. Additively manufactured hierarchical stainless steels with high strength and ductility. *Nat. Mater.* **17**, 63–71 (2017).
81. Wilkinson, A. J. & Hirsch, P. B. Electron diffraction based techniques in scanning electron microscopy of bulk materials. *Micron* **28**, 279–308 (1997).
82. Zaefferer, S. & Elhami, N.-N. Theory and application of electron channelling contrast imaging under controlled diffraction conditions. *Acta Materialia* **75**, 20–50 (2014).
83. Kong, D. et al. About metastable cellular structure in additively manufactured austenitic stainless steels. *Addit. Manuf.* **38**, 101804 (2021).
84. Godec, M., Zaefferer, S., Podgornik, B., Šinko, M. & Tchernychova, E. Quantitative multiscale correlative microstructure analysis of additive manufacturing of stainless steel 316l processed by selective laser melting. *Mater. Charact.* **160**, 110074 (2020).
85. Wang, Z., Gu, J., An, D., Liu, Y. & Song, M. Characterization of the microstructure and deformation substructure evolution in a hierarchal high-entropy alloy by correlative EBSD and ECCI. *Intermetallics* **121**, 106788 (2020).
86. Burnett, T. L. & Withers, P. J. Completing the picture through correlative characterization. *Nat. Mater.* **18**, 1041–1049 (2019).

87. Ram, F., Wright, S., Singh, S. & De Graef, M. Error analysis of the crystal orientations obtained by the dictionary approach to ebsd indexing. *Ultramicroscopy* **181**, 17–26 (2017).
88. Demers, H. et al. Three-dimensional electron microscopy simulation with the Casino Monte Carlo software. *Scanning* **33**, 135–146 (2011).
89. Gallet, J. et al. Experimental measurement of dislocation density in metallic materials: a quantitative comparison between measurement techniques (XRD, R-ECCI, HR-EBSD, TEM). *Mater. Charact.* **199**, 112842 (2023).
90. Crimp, M. A. Scanning electron microscopy imaging of dislocations in bulk materials, using electron channeling contrast. *Microsc. Res. Tech.* **69**, 374–381 (2006).
91. Zhou, B., Khosla, A., Lapedriza, A., Oliva, A. & Torralba, A. Learning deep features for discriminative localization. In *Proc. IEEE Conference on Computer Vision and Pattern Recognition*, 2921–2929 (IEEE, 2016).
92. Zeiler, M. Visualizing and understanding convolutional networks. In *Proc. Eur. Conference on Computer Vision/arXiv*, vol. 1311 (2014).
93. Selvaraju, R. R. et al. Grad-cam: Visual explanations from deep networks via gradient-based localization. In *Proc. IEEE iNternational Conference on Computer Vision*, 618–626 (IEEE, 2017).
94. Vinogradova, K., Dibrov, A. & Myers, G. Towards interpretable semantic segmentation via gradient-weighted class activation mapping (student abstract). *Proc. AAAI Conf. Artif. Intell.* 34, 13943–13944 (2020).
95. Ribeiro, M. T., Singh, S. & Guestrin, C. why should i trust you? explaining the predictions of any classifier. In *Proc. 22nd ACM SIGKDD International Conference on Knowledge Discovery and Data Mining*, 1135–1144 (SIGKDD, 2016).
96. Li, X. et al. G-lime: Statistical learning for local interpretations of deep neural networks using global priors. *Artif. Intell.* **314**, 103823 (2023).
97. Jangid, D. K. et al. Q-rbsa: high-resolution 3D ebsd map generation using an efficient quaternion transformer network. *npj Comput. Mater.* **10**, 27 (2024).
98. Jung, J. et al. Super-resolving material microstructure image via deep learning for microstructure characterization and mechanical behavior analysis. *npj Comput. Mater.* **7**, 96 (2021).
99. Calvat, M. et al. Plasticity encoding and mapping during elementary loading for accelerated mechanical properties prediction, DOI: 10.48550/ARXIV.2503.19799 (2025).
100. Paysan, P., Knothe, R., Amberg, B., Romdhani, S. & Vetter, T. A 3d face model for pose and illumination invariant face recognition. In *Proc. IEEE International Conference on Advanced Video and Signal Based Surveillance*, 296–301 (IEEE, 2009).
101. Aubry, M., Maturana, D., Efros, A. A., Russell, B. C. & Sivic, J. Seeing 3D chairs: exemplar part-based 2D-3D alignment using a large dataset of CAD models. In *Proc. IEEE Conference On Computer Vision And Pattern Recognition*, 3762–3769 (IEEE, 2014).
102. Burgess, C. P. et al. Understanding disentangling in $\beta$-vae. *arXiv preprint arXiv:1804.03599* (2018).
103. Fil, M., Mesinovic, M., Morris, M. & Wildberger, J. beta-vae reproducibility: challenges and extensions. *arXiv preprint arXiv:2112.14278* (2021).
104. Wiatrak, M., Albrecht, S. V. & Nystrom, A. Stabilizing generative adversarial networks: a survey. *arXiv preprint arXiv:1910.00927* (2019).
105. Ding, Z. & De Graef, M. Parametric simulation of electron backscatter diffraction patterns through generative models. *npj Comput. Mater.* **9**, 199 (2023).
106. Burgess, J. et al. Orientation-invariant autoencoders learn robust representations for shape profiling of cells and organelles. *Nat. Commun.* **15**, 1022 (2024).
107. McKinney, M. et al. Unsupervised multimodal fusion of in-process sensor data for advanced manufacturing process monitoring. *J. Manuf. Syst.* **78**, 271–282 (2025).
108. Hamilton, M., Zhang, Z., Hariharan, B., Snavely, N. & Freeman, W. T. Unsupervised semantic segmentation by distilling feature correspondences. *arXiv preprint arXiv:2203.08414* (2022).
109. Lu, L., Shin, Y., Su, Y. & Karniadakis, G. E. Dying ReLU and initialization: Theory and numerical examples. *arXiv preprint arXiv:1903.06733* (2019).
110. MathWorks. *Train Variational Autoencoder (VAE) to Generate Images*. MathWorks, Natick, MA (2024). Accessed: 2025-06-20.
111. Kingma, D. P. Adam: A method for stochastic optimization. *arXiv preprint arXiv:1412.6980* (2014).

## Acknowledgements

M.C., C.B., H.W, K.V. and J.C.S. are grateful for financial support from DARPA. C.B., D.A., H.P. and J.C.S. acknowledge the NSF (award #2338346) for financial support. This work was carried out in the Materials Research Laboratory Central Research Facilities, University of Illinois. Carpenter Technology is acknowledged for providing the 718 material. Morad Behandish and Adrian Lew are acknowledged for their support and leadership. Tresa Pollock, McLean Echlin and James Lamb are acknowledged for their support on the EBSD sharpness calculations. Marat Latypov, Marie Charpagne, and Florian Strub are gratefully acknowledged for their support and insightful discussions.

## Author contributions

M. C.: Conceptualization, Data curation, Formal analysis, Investigation, Methodology, Writing – original draft, Writing – review & editing. C. B.: Conceptualization, Data curation, Formal analysis, Investigation, Methodology, Writing – original draft, Writing – review & editing. D. A.: Data curation, Verification, Writing – review & editing. H. P.: Verification, Writing – review & editing. H. W.: Resources. K. V.: Writing – review & editing. J. C. S.: Conceptualization, Funding acquisition, Methodology, Project administration, Resources, Supervision, Writing – original draft, Writing – review & editing.

## Competing interests

The authors declare no competing interests.

## Additional information

**Correspondence** and requests for materials should be addressed to Mathieu Calvat or J. C. Stinville.